# Giant proton transmembrane transport through sulfophenylated graphene in a direct methanol fuel cell

Weizhe Zhang[1]*, Max Makurat[1], Xue Liu[1,2], Xiaofang Kang[1], Xiaoting Liu[3,4], Yanglizhi Li[3,4], Thomas J.F. Kock[1], Christopher Leist[5], Clément Maheu[6], Hikmet Sezen[6], Lin Jiang[1], Dario Calvani[1], Andy Jiao[1], Ismail Eren[7], Francesco Buda[1], Agnieszka Kuc[7], Thomas Heine[8,9], Haoyuan Qi[5], Xinliang Feng[10,11], Jan P. Hofmann[6], Ute Kaiser[5], Luzhao Sun[3,4], Zhongfan Liu[3,4], Grégory F. Schneider[1]*

[1] *Leiden Institute of Chemistry, Faculty of Science, Leiden University, Einsteinweg 55, 2333CC Leiden, The Netherlands.*

[2] *now: State Key Laboratory for Mechanical Behavior of Materials, Xi'an Jiaotong University, 710049 Xi'an, China.*

[3] *Center for Nanochemistry, Beijing Science and Engineering Center for Nanocarbons, Beijing National Laboratory for Molecular Sciences, College of Chemistry and Molecular Engineering, Academy for Advanced Interdisciplinary Studies, Peking University, Beijing, China.*

[4] *Beijing Graphene Institute (BGI), Beijing, China.*

[5] *Central Facility of Electron Microscopy, Electron Microscopy Group of Materials Science, Ulm University, Ulm, Germany.*

[6] *Surface Science Laboratory, Department of Materials and Earth Sciences, Technical University of Darmstadt, Otto-Berndt-Strasse 3, 64287 Darmstadt, Germany.*

[7] *Helmholtz-Zentrum Dresden-Rossendorf, Abteilung Ressourcenökologie, Permoserstrasse 15, 04318 Leipzig, Germany.*

[8] *Theoretical Chemistry, Technical University Dresden, 01062 Dresden, Germany.*

[9] *Yonsei University and ibs-cnm, Seodaemun-gu, Seoul 120-749, Republic of Korea*

[10] *Center for Advancing Electronics Dresden & Faculty of Chemistry and Food Chemistry, Technische Universität Dresden, Dresden, Germany.*

[11] *Max Planck Institute of Microstructure Physics, 06120 Halle (Saale), Germany.*

* to whom correspondence should be addressed:

w.zhang@lic.leidenuniv.nl, g.f.schneider@chem.leidenuniv.nl

**Abstract.** *(128 words)*

An ideal proton exchange membrane should only permeate protons and be leak tight for fuels. Graphene is impermeable to water and poorly conducting to protons. Here, we chemically functionalized monolayer graphene to install sulfophenylated $sp^3$ dislocations by diazotization. Selective to protons, transmembrane areal conductances are up to ~50 S cm$^{-2}$, which is ~5000 fold higher than in pristine graphene. Mounted in a direct methanol fuel cell, sulfophenylated graphene resulted in power densities up to 1.6 W mg$^{-1}$ or 123 mW cm$^{-2}$ under standard cell operation (60 °C), a value ~two-fold larger than micron-thick films of Nafion 117. The combination of $sp^3$ dislocations and polar groups, therefore, allow the creation of hydrophilic ion paths through graphene and unveils a novel route to rationalize transmembrane hydron transport through 2D materials.

**One-sentence summary.** Opening proton channels in graphene by sulfonation.

**Main text.**

Next to long-term stability an ideal and optimized proton exchange membrane needs to fulfil two main criteria: proton permeability and selectivity. Within methanol fuel cells, the first ensures a high power density, while the second prevents fuel cross-over between the electrodes, which deteriorates catalyst performance and, thereby, drastically lowers performance. Pristine graphene(*1*) already fulfils these two criteria, partly as the graphene basal plane is impermeable to water and other molecules(*2*), and exhibits a certain degree of proton conductivity(*3*). In polymer membrane development, however, proton conductivity and selectivity are antagonistic with respect to their performance(*4*). Long channel length in state-of-the-art membranes such as Nafion 117, is, therefore, a prerequisite to obtaining proton selectivity, at the cost of an additional ionic resistance path through such long channels. With channel diameters of several nanometers(*5, 6*), being ~10 times larger than the size of hydrated protons and molecular fuels, fuel crossover through the membrane usually hinders optimal membrane performance.

Protons(*3*), as one form of hydrons(*7*), can translocate through graphene and other 2D materials such as hexagonal boron nitride (*h*-BN)(*8*) and 2D mica(*9*) at room temperature, making two-dimensional (2D) materials promising candidates for proton-exchange membrane applications(*10*). Beyond monolayer 2D materials, proton transport has also been demonstrated using 2D laminates (*i.e.*, 2D crystals assembled in a layered structure, made from graphene flakes(*11, 12*), and other 2D nanosheets(*13, 14*), and covalent organic frameworks(*15*)), making them suitable for membrane-based applications. Remarkably, functionalizing 2D materials and reducing interlayer distances down to sub-nanometre dimensions yielded membranes with high selectivity, as well as opportunities to observe ion transport under highly confined spaces(*16, 17*). However, the ion transport distance in these systems is currently, particularly in laminated structures(*18*), beyond 'one atom', thus yielding additional parallel resistances. Protons were also shown not to permeate through defect-free regions of graphene(*19*), which demonstrates the need for proton-selective pathways. Open questions focus on the role of defects and strain(*20*), and – we believe with our work – also the role of out-of-plane polar $sp^3$ dislocations installed onto the graphene basal plane.

To open pores enabling transmembrane ion transport in graphene(*21*) and other 2D materials, such as $MoS_2$(*22, 23*), methods have been established using electron beams, ion bombardment, and plasma exposure(*24-31*). Ion bombardment allows obtaining a high pore density and small pores ($>10^{14}$ cm$^{-2}$)(*25*), while electron beam sculpting, suffers from a lower control over the resulting pore size distribution, and chemistry. Pore and pore-like defects have also been realized via bottom-up approaches including chemical vapor deposition (CVD) growth of amorphous carbon(*32, 33*), using 2D polymers(*34, 35*), and graphyne(*36*). These bottom-up strategies benefit from remarkable scalability in combination with the possibility of controlling the size, shape, and chemistry of the pore, unfortunately yet with pores that are larger than the diameter of a hydrated proton. In fact, localized defects in graphene, such as seven- and higher-member rings together with small lattice disorders, such as $sp^3$ defects, are promising approaches towards increased proton conductivity, while keeping the basal plane impermeable to other substances(*37-39*).

Here, we introduce $sp^3$ lattice dislocations paired with sulfophenyl groups on the basal plane of single-layer graphene via a diazotization strategy(*40*) using 4-sulfobenzenediazonium (4-SBD) as a reactant (Fig. 1A). The functionalization converts $sp^2$ carbons from graphene into functional $sp^3$ dislocations carrying a hard charge (from the sulfonate; $pK_a$ ~ -2.8 in water) and a resulting out-of-plane dipole. A delocalized electron from the graphene lattice reacts with the 4-SBD cation leading to a radical formation as $N_2$ is expelled as a leaving group. The resulting aryl radical reacts with a $sp^2$ carbon atom on the graphene basal plane, which is converted to $sp^3$ upon grafting. Meanwhile, side reactions might also occur: 4-SBD can react with an already grafted aryl group (at the ortho position) either via a radical or azo-coupling mechanism, both of which might result in oligomerization(*41*) of the aryl species on the graphene basal plane.

To eliminate the influence of intrinsic defects and to understand proton transport through the sole sulfophenylated sites on graphene (now abbreviated $SO_3^-$ graphene), we first used single crystals of mechanically exfoliated graphene. We measured the proton current through free-standing graphene supported over a 1μm diameter circular aperture in a ~30 nm thick custom-made SiN chip (Fig. 1B). Next, we functionalized exfoliated graphene with 4-SBD (1 mg ml$^{-1}$ in 0.1 M HCl). Upon functionalization the areal ionic conductance in 0.1 M HCl increased up to ~50 S cm$^{-2}$ after 84 hours of treatment (Fig. 1C). Respectively, up to 7 S cm$^{-2}$ in 0.1M KCl, showing a significant selectivity for proton compare to potassium ions. For treatments longer than 84 hours, the ionic current decreased in both HCl and KCl electrolytes, which we attribute to 4-SBD oligomerization occurring at anchored aryl groups resulting in an additional resistance (phase II in Fig. 1C). Additionally, and as expected, larger multivalent cations such as $Cu^{2+}$, $Ca^{2+}$, and $Fe^{3+}$, in chloride electrolytes resulted in lower overall conductances compared to protons (see inset of Fig. 1C). For the three measured devices, the maximum areal conductance in 0.1M HCl reached, respectively, 50.0 S cm$^{-2}$ (Fig. S14; device 1), 49.2 S cm$^{-2}$ (Fig. S17, device 2), and 24.2 S cm$^{-2}$ (Fig. S19, device 3) in 0.1 M HCl. These conductance values are up to ~5000 times higher than those of pristine graphene (*i.e.*, 0.1 S cm$^{-2}$ for device 1; 0.008 S cm$^{-2}$ for device 2; 0.1 S cm$^{-2}$ for device 3) in HCl 0.1M before starting the 4-SBD treatment (i.e., at 0h in Fig 1C). The giant measured areal proton conductance of 50.0 S cm$^{-2}$ for $SO_3^-$ graphene corresponds to a Gibbs free energy barrier ΔG ~0.42 eV following the Nernst-Einstein equation (Fig. 1D). This energy is similar to that found for the proton release from the sulfophenylated functionality into the water bulk, which turned out to be the rate-limiting step for the proton permeation through a graphene nanopore.[1] For a sulfophenylated sp³ dislocation on a pristine graphene monolayer, the process could proceed as follows: first, the proton is captured from the water bulk by the $SO_3^-$ functionality and then released with a similar activation barrier $\Delta G \sim 0.47$ eV (see Supporting Information, 2.2). Next, via the Grotthuss mechanism, the proton would reach the graphene basal plane and tunnel to the other side of the graphene. The energy barrier of the proton quantum tunneling is estimated to be ~0.4-3 eV(*42*), which could be comparable to the barrier displayed in the proton release from the sulfophenylated functionality. In practice, the sulfophenylated functionality can play the role of a proton shuttle from the water bulk to the proximity of the sp³ dislocation on the graphene basal plane requesting an activation energy down to ~0.4-0.5 eV (Fig. S20). [1] Such value is in agreement with an efficient proton permeability and selectivity through a graphene-based membrane, giving the measured areal proton conductance of ~50 S cm⁻² and extraordinary selectivity for protons (Fig. 1D).

---

[1] See additional review material. Dario Calvani et al, submitted (March 2023)

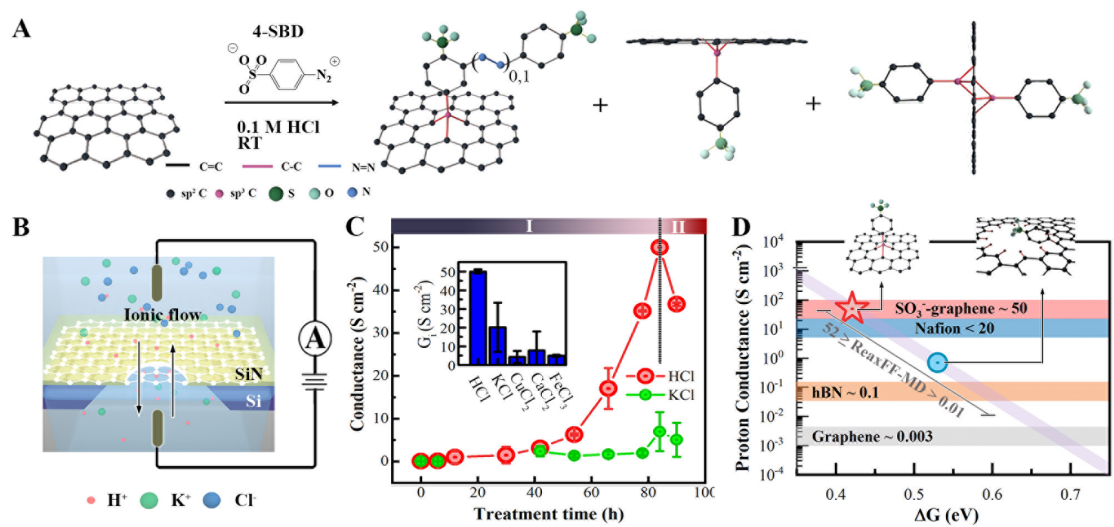

**Fig. 1 Proton-selective sub-nanometer pathway in exfoliated single-layer monocrystalline graphene by diazotization with 4-sulfobenzenediazonium tetrafluoroborate (4-SBD). (A)** Illustration depicting the functionalization of

graphene with 4-SBD. Besides the direct coupling of the sulfophenylated radical to the graphene basal plane, which converts $sp^2$ carbon to $sp^3$, 4-SBD could also link to the ortho position of a previously grafted benzene ring via a radical coupling or an azo coupling. **(B)** Illustration of the ionic current measurements setup. An exfoliated graphene flake is transferred on a Si/SiN chip with a 1 μm in diameter circular aperture in a 30nm thick SiN free-standing membrane (details on the chip fabrication procedure can be found in the supplementary information, section 2). The as-prepared chip was then mounted in a flow cell containing two reservoirs containing electrolyte solutions (HCl, KCl, $CuCl_2$, $CaCl_2$, and $FeCl_3$). The transmembrane ionic current was measured by applying a 100 mV bias between two Ag/AgCl immersed on both sides of the graphene layer. From current-voltage curves (I-V, Fig. S14-S17) the areal conductance was determined. **(C)** Areal conductance (at 100 mV) measured in 0.1M HCl and 0.1M KCl as a function of 4-SBD reaction times ranging from 0 to 96 h. Inset: Normalized areal conductance measured in 0.1M electrolytes for HCl, KCl, $CuCl_2$, $CaCl_2$, and $FeCl_3$. For the discussion of selectivity, conductance of specific chloride salt were normalized to HCl by dividing the measured conductivity by bulk conductivities for each salt (details in section 2 of the supplementary information). The error bar is the standard deviation between devices 1 and 2. **(D)** Linear correlation plot between the estimated proton conductance (S $cm^{-2}$) in logarithmic scale and Gibbs free energy barrier, ΔG (in eV) for the proton diffusion using the Nernst-Einstein equation(*43*). The grey and orange bars highlight the areal proton conductance of graphene and hexagonal boron-nitride (hBN) reported in ref. (*3*). The blue bar represents the calculated areal proton conductance of Nafion assuming a thickness of 20 μm with the upper conductance limit reported in ref. (*44*). The red bar represents the measured areal proton conductance for $SO_3^-$ graphene in our setup (*i.e.*, from 24.4 to 50.0 S cm$^{-2}$). The red star represents the proton conductance of device 1 after 84h 4-SBD treatment and corresponds to the DOE 2025 target (*45*). The blue circle is the average estimated Gibbs free energy barrier and the diagonal grey bar is the corresponding range of proton conductance obtained by ReaxFF-MD simulations of a graphene nanopore (1 nm diameter) functionalized with a single sulfophenylated group.[1]

Next, we prepared $SO_3^-$-graphene as a proton-selective membrane in a direct methanol fuel cell (DMFC). To apply $SO_3^-$-graphene in a DMFC (Fig. 2A), mechanically exfoliated graphene could not be used as it is limited to micron-sized samples. We, therefore, replaced exfoliated graphene with centimeter-size single-crystal graphene films, which were seamlessly stitched into well-aligned domains(*46*). Fig. 2B shows scanning electron microscopy (SEM) images of >200 μm wide CVD graphene single-crystals with minimized grain boundaries and intrinsic defects grown on a copper foil. Multilayer patches were negligibly observed in the samples. We then transferred CVD graphene to a quantifoil TEM grid and incubated graphene with 4-SBD for up to 6 days (1 mg ml$^{-1}$ 4-SBD in 0.1 M HCl) and characterized $SO_3^-$-graphene by atomic force microscopy (AFM, Fig. 2C) and high-resolution transmission electron microscopy (HRTEM, Fig. 2D). From the AFM images, the arithmetic average (Ra) and root mean square roughness (RMS) increased by 0.5nm within the first 24 h of incubation (Fig.

2C). After 24 h, only a minimal change in surface roughness occurred (Fig. 2C and Fig. S28). However, from the error image channels (for a spatially better-resolved image) a clear change in morphology occurred for longer treatment times: graphene surface features such as folds resulting from the transfer became decreasingly visible while denser agglomerates appeared upon longer 4-SBD treatments. Pristine graphene is therefore covered by 4-SBD already after 24 h after which the 4-SBD layer becomes thicker and more homogenous. Oligomerization, which is commonly observed during diazonium treatment(*40*) and here on AFM images, likely leads to the observed additional resistance that protons encounter resulting in the significant decrease in proton conductance after 84h of treatment as shown in Fig. 1C.

Importantly, at the nanometer scale, HRTEM images of graphene after 4-day and 6-day of SBD treatment did not show the apparition of large >1 nm pores (Fig. 1D). Instead, the large defect-free graphene area suggests that individual grafting sites could not be observed on the monolayer regions (Fig. 1D). The spontaneous oligomerization of SBD, giving rise to agglomerates visible in AFM, were therefore not distinguishable by HRTEM from common hydrocarbon contaminations resulting from the growth of CVD graphene (supplementary material, Fig. S29-S31).

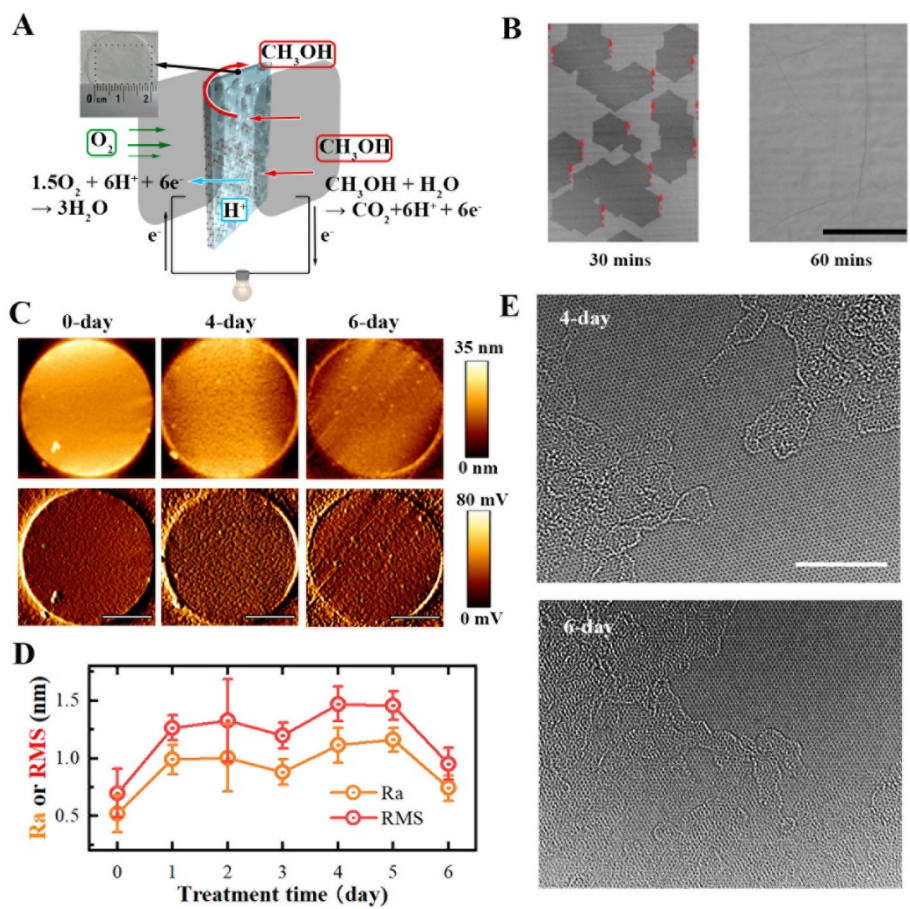

**Fig. 2 $SO_3^-$-graphene in a direct methanol fuel cell (DMFC), growth of single-crystalline chemical vapor deposition (CVD) graphene, and characterization by scanning electron microscopy (SEM), atomic force microscopy (AFM), and high-resolution transmission electron microscopy (HRTEM).** **(A)** Illustration of a DMFC, redox reactions involved in the conversion of methanol to carbon dioxide, and photograph of CVD graphene transferred on a polycarbonate support. **(B)** SEM images of hexagonal graphene single crystals after 30 mins of growth on a copper foil (left) to a fully-covered and confluent monolayer after one hour of growth (right). Scale bar: 200 μm. (**C**) AFM images of pristine free-standing single-crystalline CVD graphene transferred over a holey quantifoil TEM grid and after 4 and 6 days of 4-SBD treatment. The graphene was transferred to a TEM grid using a polymer-free transfer method by depositing the TEM grid on the graphene side of the copper foil and floating the TEM grid and graphene/copper stack on a 0.5M ammonium persulfate aqueous copper-etching solution. Scale bar: 500 nm. (**D**) Arithmetic roughness average (Ra) and root means square average (RMS) of graphene and 4-SBD treated $SO_3^-$ treated graphene derived from AFM image analysis. (**E**) HRTEM image of graphene after 4-day and 6-day 4-SBD treatment. Scale bar: 5 nm.

Next, we studied 4-SBD reactivity with graphene covered with a spin-coated Nafion layer (Fig. 3Ai). The choice of Nafion (instead of PMMA) as a support polymer is because Nafion will be used as a support film for graphene in the DMFC setup compromising both the mechanical stability of graphene and being freestanding in water. For both graphene carried by Nafion (Fig. 3Aiii) and bare graphene on $SiO_2$/Si (Fig. S25), Raman spectroscopy showed an increasing D peak for increasing 4-SBD treatment times and $I_D/I_G$ reached a plateau indicating a maximum reactivity after 4 days of treatment. The increasing value of the D peak intensity ratio (~1350 cm$^{-1}$) over the G peak (~1580 cm$^{-1}$) ($I_D/I_G$) indicates the formation of defects upon 4-SBD treatment, primarily the introduction of sp$^3$ dislocations(*47*). Moreover, the intensity ratio of the 2D peak (~2700 cm$^{-1}$) over the G peak ($0.6 < I_{2D}/I_G < 0.8$) indicates no major changes in the quality and number of layers of graphene, although the value is not 2.0 in the case of polymer (Nafion) coated graphene(*48*)(see supplementary information Fig. S25, $I_{2D}/I_G > 2.0$ for untreated graphene, on $SiO_2$/Si, and after 5 days treatment $I_{2D}/I_G = 0.6$).

Additionally to Raman, we used X-ray photoelectron spectroscopy (XPS) to study the sulphur content and hybridization of the carbon atoms on the lattice upon 4-SBD functionalization. Fig. 3B shows XPS core level spectra (C1s, N1s, and S2p) of pristine graphene and after 2 days, 4 days, and 6 days of treatment. The fitting of the C1s spectra leads to four distinct species including the characteristic C sp$^2$ (BE≈284.5 eV) and C sp$^3$ (BE≈285.3 eV) of graphene. Moreover, semi-quantitative analyses were also conducted to determine the chemical composition of the grafted sites (Table S1). The sp$^3$C/sp$^2$C ratio reveals the state of sp$^3$/sp$^2$ conversion of graphene upon 4-SBD treatment. Relative ratios of -SO$_3^-$ and -N=N- indicate the number of sulfophenyl groups on the graphene surface. For ease of comparison, we plotted sp$^2$C/sp$^3$C and -SO$_3^-$, -N=N- in Fig. 3C. The sp$^3$C/sp$^2$C ratio increases from 0.2 for pristine graphene up to 1.0 after 2-day treatment. After 6-day treatment, the sp$^3$C/sp$^2$C ratio dropped to 0.3. Considering that a benzene ring contains six sp$^2$C, we assume that the drop of sp$^3$/sp$^2$ ratio after 4-day and 6-day treatment is due to the addition of 4-SBD groups via an oligomerization mechanism instead of grafting to graphene directly. Accordingly, the relative ratio of -SO$_3^-$ increases from 0 for pristine graphene to 5.2 after 4 days of treatment and further increases to 6.4 after 6-day of treatment. The relative ratio of -N=N- shows a similar trend to SO$_3^-$, which also supports simultaneous azo coupling of sulfophenyl groups, in agreement with AFM. According to the Raman results, however, the intensity of the D peak plateaued after 4 days of treatment, suggesting that only oligomerization (coupling between 4-SBD and grafted sulfonatophenyl groups) occurred for treatment times longer than 4 days.

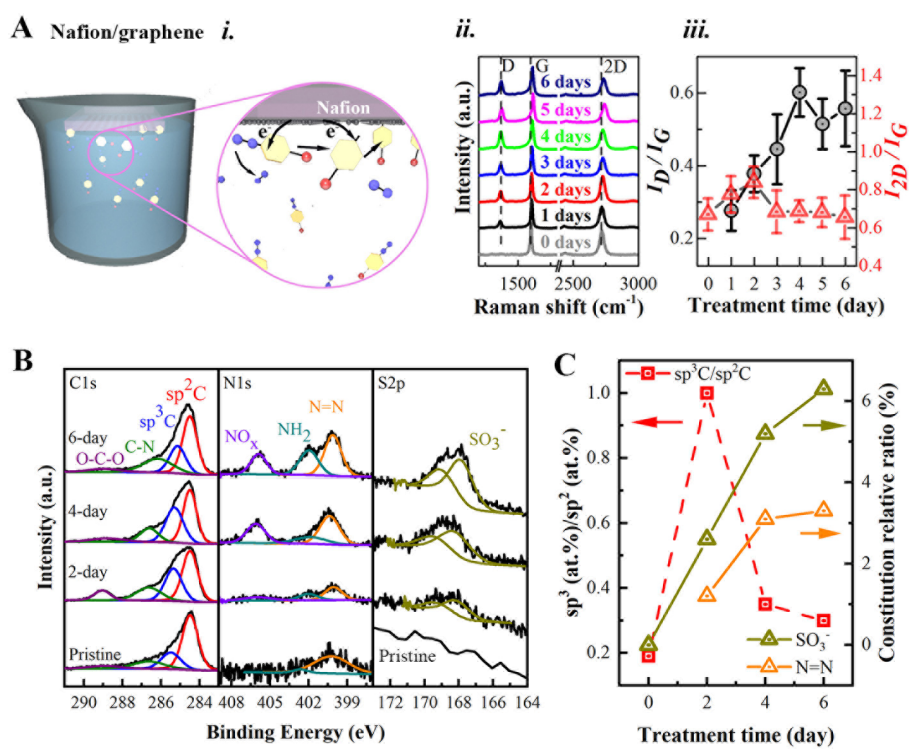

**Fig. 3 Raman spectroscopy and X-ray photoelectron spectroscopy (XPS) analysis of $SO_3^-$-graphene.** (**A**) i) Illustration of graphene coated with Nafion floating on and reacting with a solution of 4-SBD (1 mg/ml SBD in aqueous 0.1 M HCl). ii) Raw Raman spectra for graphene on Nafion transferred on a $SiO_2$/Si substrate after 0-6 days of floating incubation with 4-SBD. Raman spectra were recorded over a 10×10 μm area using a 457 nm laser set to 1.5 mW power to avoid any laser-induced damage to the $SO_3^-$-graphene. We normalized the spectra by the intensity of the G peak to facilitate a comparison of the D and 2D peaks. iii) Plot of $I_D/I_G$ and $I_{2D}/I_G$ versus 4-SBD treatment time derived from panel Aii). (**B**) C1s, N1s, and S2p XPS core level spectra of pristine graphene and graphene after 2-day, 4-day, and 6-day of 4-SBD treatment on $SiO_2$/Si substrate with fitted components. (**C**) Composition relative ratios as a function of 4-SBD reaction time. The ratios are derived for the specific component from the overall chemical bonds characterized by XPS. $sp^3C/sp^2C$ ratio, N=N and $SO_3^-$ were plotted

together for comparison with $sp^3C/sp^2C$ ratio linked to the left and $N=N/SO_3^-$ linked to the right y-axis.

Next, we tested $SO_3^-$-graphene as a membrane in a DMFC. We measured three independent batches of $SO_3^-$-graphene after each day of treatment (*i.e.*, after 0, 1, 2, 3, 4, 5, and 6 days; 21 samples in total) and determined maximum power densities, membrane conductance, and methanol crossover. Fig. 4A shows the power-current (P-V) and voltage-current (V-I) plots for pristine CVD graphene, Nafion, and $SO_3^-$-graphene at 60 °C. Pristine graphene showed a maximum power density of 25.8 ±15.4 mW cm$^{-2}$, at 212 mV, which is about half (~ 45%) compared to Nafion 117 and a proton conductance of 2.4 S cm$^{-2}$. These results can be attributed to defects resulting from the membrane electrode assembly (MEA) including CVD graphene(*20*). It is notable that, however, for $SO_3^-$-graphene after 4 days of reaction time with 4-SBD, the maximum power density increased to 109.8 ±14.8 mW cm$^{-2}$ (*i.e.*, ~2 times higher than Nafion). The power output did not increase further with a prolonged 4-SBD reaction time. Instead, it decreased to 44.9 ±17.3 mW cm$^{-2}$ after a 6-day treatment, again in line with our hypothesis of the oligomerization of the sulfophenyls after 4 days of treatment. Subsequently, we monitored the power density for a range of operating temperatures ranging from room temperature (rt) to 70 °C (Fig. 4B). In general, an increase in temperature leads to an elevated catalytic activity as well as an increased diffusion rate of both protons and methanol across the membrane. As expected, the power density for 4-day treated $SO_3^-$-graphene increased linearly with the temperature rising from 20 °C to 70 °C (Fig. 4B). In contrast, for pristine graphene and 6-day treated graphene, the maximum power densities did not show a linear increase with temperature, and even decreased for operation temperatures above 50 °C. We attribute this observation to a higher methanol crossover rate at higher temperatures suggesting the importance of an optimal 4-SBD treatment time of 4 days. In fact, in the pristine sample, proton transport occurs primarily through inherent defects, which have poor proton selectivity and we presume therefore more methanol crossover. Similarly, for the oligomerized samples, the proton selectivity sites are proven to be blocked, resulting in relatively low proton conductance. As a result, inherent defects also facilitate the methanol crossover.

To establish a correlation between areal proton conductance and DMFC performance (*i.e.*, power density; and in the next section also fuel crossover rates), we first investigated how the membrane resistance of $SO_3^-$-graphene correlates with the operating temperature of the fuel cell. Fig. 4C shows the Nyquist plots obtained by electrochemical impedance spectroscopy (EIS), in which the intercept of the x-axis indicates the overall resistance of the membrane. Remarkably, the lowest resistance was also achieved after 4 days of treatment with 4-SBD (*i.e.*, R = 0.59 Ω, which is ~ 30% lower than 6-day treated graphene and 2.5-fold lower than Nafion 117). The higher resistance after 6 days of reaction with 4-SBD also suggests a hinder of the proton transport pathway due to oligomerization. In ionic measurements (Fig. 1), however, the maximum areal conductance was reached earlier in time (*i.e.*, 84h, 3.5-day). This bias in the time to reach a maximum conductance (respectively, 84h for exfoliated graphene, and 4 days, *i.e.* ~96h for CVD graphene) is likely due to a difference in reactivity of the graphene in both ion transport and DMFC setups. Importantly, pristine CVD graphene shows relatively larger proton conductance, normalized to the exposed CVD graphene area, compared to the exfoliated graphene (6.9 ±1.1 S cm$^{-2}$ *vs.* 0.008-0.1 S cm$^{-2}$ respectively), which originates from defects in the form of pinholes or cracks in CVD graphene (*20, 49*). For SO$_3^-$-graphene, similarly, the areal conductance observed after 4 days of reaction time is 4.9 ±0.4 S cm$^{-2}$ per electrode area. As opposed to ionic measurements, in the DMFC, electrodes cover the entire SO$_3^-$-graphene membrane while graphene is only free-standing over 16% compared to the electrode area (*i.e.* only ~ 16% of the polycarbonate membrane area is open with free-standing graphene). Therefore, the equivalent conductance per graphene area in the DMFC is 30.9 ±2.3 S cm$^{-2}$, a value ~19.1 S cm$^{-2}$ lower than exfoliated graphene (Fig. 1). The conductance values are, therefore, in good agreement since only one side of the graphene could be functionalized with SBD in the DMFC setup[2].

Additionally, the observed high proton selectivity among hydrated cations (Fig. 1) is also reflected by a low methanol crossover rate, particularly when comparing power densities in *P-I* curves for 1 M and 5 M methanol (Fig. 4E). After a switch from 1 M to 5 M methanol, the maximum power density with Nafion 117 dropped by ~53%. The drop in maximum power density, switching methanol from 1 M to 5 M, was reduced to ~10% after 4-days of reaction time with 4-SBD (Fig. 4F). The sulfonation of graphene using 4-SBD, therefore, not only allows an efficient selectivity towards proton transport but also prevents fuels (here methanol) from crossing the membrane in a DMFC.

---

[2] After 4 days 4-SBD treatment for **single side**, the increase in conductance of **CVD graphene** is 30.9 - 6.9 = 24.0 S cm$^{-2}$. After 84h 4-SBD treatment for **both sides**, the increase in conductance of **exfoliated graphene** is 50.0 - 0.1 = 49.9 S cm$^{-2}$.

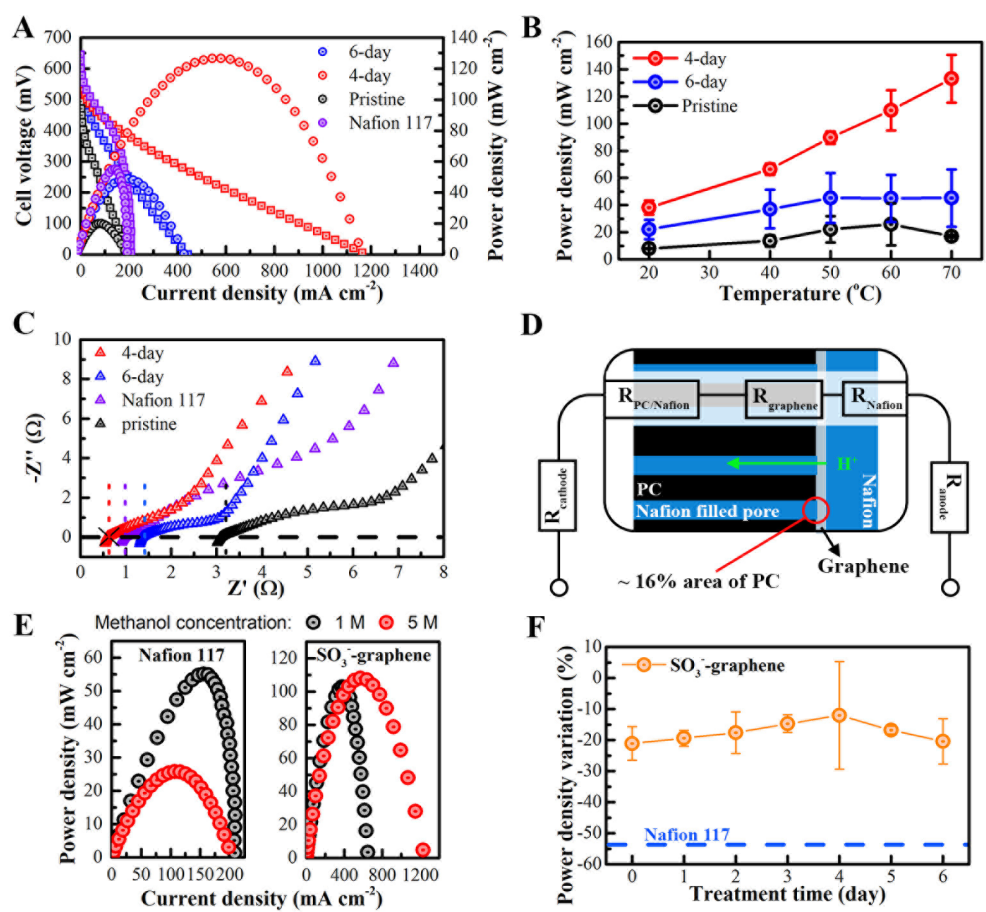

**Fig. 4. DMFC performance with CVD graphene and SO$_3^-$-graphene membranes.** **(A)** Power-current density (*P-I*) and voltage-current density (*V-I*) curves for Nafion 117, pristine graphene, 4-day, and 6-day 4-SBD treated SO$_3^-$-graphene. The cell voltage *(V)* and the power density *(P)* as a function of current density *(I)* were measured in a 1 M methanol/water solution at 60 °C. **(B)** Maximum power density as a function of temperature. **(C)** Nyquist plots of electrochemical impedance spectroscopy (EIS) measured at the open circuit voltage. **(D)** Equivalent circuit of the DMFC with graphene and SO$_3^-$-graphene membranes. The proton conductive pathway is constituted by Nafion-filled pores (~16% of PC) in the PC membrane, graphene, and a spin-coated Nafion ionomer layer. **(E)** *P-I* curves of fuel cell with 1 and 5 M methanol fuel for Nafion 117 and $SO_3^-$-graphene (4-day treatment) under 60 °C. **(F)** Variation of maximum power density upon operating the DMFC in 1 and 5 M methanol respectively. The blue dashed line corresponds to the variation of the power density of a DMFC with only Nafion 117 when switching from 1 M methanol to 5 M methanol.

To conclude, the functionalization of graphene (exfoliated and CVD) with an out-of-plane polar group and an $sp^3$ dislocation by diazotization with a sulfophenyl radical opened an atomically thin transmembrane proton selective path which yielded highly selective proton conductivity (~50 S cm$^{-2}$, room temperature) and also enabled application in a direct methanol fuel cell delivering a power density more than twice higher than its Nafion counterpart according to the electrode area (127 mW cm$^{-2}$ at 60 °C). This finding calls for more extensive views and understandings of the crucial reactions and mechanisms at the origin of such giant conductance value, among the highest reported now for 2D materials (*50*) and importantly within the 2025 DOE target for the proton conductance of a fuel cell under operation conditions of 90 °C and 25 kPa water partial pressure (*45*). Particular attention is required to understand the proton transport through $sp^3$ distorted graphene, including theoretical insights on how the proton tunneling barrier could be further tuned, and the importance of the geometry and polarity of the functional distortions. Practically, improving the $sp^3$ density might further increase the proton conductance through functionalized graphene. On a fundamental aspect, polarizing a 2D membrane may become increasingly important in controlling with precision the translocation-transport of ions, also for example in more complex two-dimensional polymer architectures(*51, 52*).

**References.**

**Acknowledgments.**

This research was supported by the European Research Council under the European Union's Seventh Framework Programme (FP/2007-2013)/ERC Grant Agreement no. 335879 project acronym "Biographene" and the Netherlands Organization for Scientific Research (NWO) VIDI 723.013.007, awarded to G.F. Schneider. Z.F. Liu, L.Z. Sun and X.T. Liu acknowledge the support from the National Natural Science Foundation of China (NSFC, No. T2188101). C. Maheu acknowledges funding by the German Research Foundation (DFG) under project no. 423746744. The authors acknowledge Hans van den Elst for carrying out the HRMS measurement. The authors thank Dr. Dima Filippov and Hugo Schellevis for helpful discussions.

**Author contributions.**

W.Z. discovered that diazotization of graphene with 4-SBD yields a giant increase of transmembrane areal proton conductance with selectivity to proton. W.Z. prepared all the graphene samples for proton transport, characterizations, and fuel cell tests, alongside the manuscript preparation. Fuel cell tests were conducted by W.Z. M.M. synthesized the 4-SBD compound and characterized it by NMR and analyzed the HRMS data. M.M. designed and carried out the AFM experiments as well as analyzed the AFM data. X.L. (Xue) proposed the idea of sulfonating graphene to form a 2D proton exchange membrane with the reported 4-SBD compound. X.L. (Xue) contributed to measuring ion transport in the very first device presented in the supplementary information. X.K helped with the exfoliated graphene sample preparation and followed-up study of ion transport. X.L. (Xiaoting), Y.L., L.S., and Z.L. synthesized the single-crystalline CVD graphene for fuel cell tests. T.J.F.K. helped with the sample preparation of graphene on TEM grids and early-stage study with TEM. A. J. designed a reactivity test for 4-SBD. C.L., H.Q., X.F., and U.K. contributed to the HRTEM characterization and discussion. C.M. and J.P.H. contributed to the XPS characterization and discussion. D.C, F.D., I.E., A.K., and T.H. contributed to the discussions and calculations to support using theoretical models of the areal conductances measured experimentally. This work was under the supervision of G.F.S. who led all the experimental progress, discussions, and the elaboration of the manuscript draft. All co-authors contributed to the writing and editing of this manuscript.

**Competing interests.**

The authors declare that they have no competing interests.

**Data and materials availability.**

All data are available in the main text or in the supplementary materials.

**Supplementary Materials**

Materials and Methods

Supplementary Text

Fig. S1 to S36

Tables S1 to S2

References (*1–35*)

# Supplementary material

*for*

# Giant proton transmembrane transport through sulfophenylated graphene in a direct methanol fuel cell

Weizhe Zhang[1]*, Max Makurat[1], Xue Liu[1,2], Xiaofang Kang[1], Xiaoting Liu[3,4], Yanglizhi Li[3,4], Thomas J.F. Kock[1], Christopher Leist[5], Clément Maheu[6], Hikmet Sezen[6], Lin Jiang[1], Dario Calvani[1], Andy Jiao[1], Ismail Eren[7], Francesco Buda[1], Agnieszka Kuc[7], Thomas Heine[8,9], Haoyuan Qi[5], Xinliang Feng[10,11], Jan P. Hofmann[6], Ute Kaiser[5], Luzhao Sun[3,4], Zhongfan Liu[3,4], Grégory F. Schneider[1]*

[1] *Leiden Institute of Chemistry, Faculty of Science, Leiden University, Einsteinweg 55, 2333CC Leiden, The Netherlands.*

[2] *now: State Key Laboratory for Mechanical Behavior of Materials, Xi'an Jiaotong University, 710049 Xi'an, China.*

[3] *Center for Nanochemistry, Beijing Science and Engineering Center for Nanocarbons, Beijing National Laboratory for Molecular Sciences, College of Chemistry and Molecular Engineering, Academy for Advanced Interdisciplinary Studies, Peking University, Beijing, China.*

[4] *Beijing Graphene Institute (BGI), Beijing, China.*

[5] *Central Facility of Electron Microscopy, Electron Microscopy Group of Materials Science, Ulm University, Ulm, Germany.*

[6] *Surface Science Laboratory, Department of Materials and Earth Sciences, Technical University of Darmstadt, Otto-Berndt-Strasse 3, 64287 Darmstadt, Germany.*

[7] *Helmholtz-Zentrum Dresden-Rossendorf, Abteilung Ressourcenökologie, Permoserstrasse 15, 04318 Leipzig, Germany.*

[8] *Theoretical Chemistry, Technical University Dresden, 01062 Dresden, Germany.*

[9] *Yonsei University and ibs-cnm, Seodaemun-gu, Seoul 120-749, Republic of Korea*

[10] *Center for Advancing Electronics Dresden & Faculty of Chemistry and Food Chemistry,*

*Technische Universität Dresden, Dresden, Germany.*

[11] *Max Planck Institute of Microstructure Physics, 06120 Halle (Saale), Germany.*

\* to whom correspondence should be addressed: w.zhang@lic.leidenuniv.nl,

g.f.schneider@chem.leidenuniv.nl

**This PDF file includes:**

1. 4-sulfonatobenzenediazonium (4-SBD) synthesis.
2. Proton (ions) conductance through exfoliated graphene before and after 4-SBD treatment.
3. Single-crystalline graphene growth and characterization.
4. Characterizations of single-crystalline graphene after 4-SBD treatment (Raman spectroscopy, XPS, AFM, and HRTEM).
5. Membrane electrode assembly (MEA) fabrication and direct methanol fuel cell (DMFC) operation.
6. Methanol crossover and fuel efficiency estimation.
7. References.

and

**Figures S1 to S36 and Tables S1 to S2.**

# 1. 4-sulfonatobenzendiazoium (4-SBD) synthesis

All solvents and reagents were obtained from commercial sources (Sigma Aldrich and Merck) and used without further purification unless stated otherwise in the respective method section. $^1$H, $^{13}$C, and $^{19}$F spectra were recorded on a Bruker AV 400 MHz spectrometer and the chemical shifts are reported in ppm (δ) with respect to deuterated solvents as an internal standard. The data is reported as chemical shift (δ), multiplicity (d = doublet), and coupling constants $J$ are reported in Hz. High-resolution mass spectra were recorded by direct injection (2 µL of a 2 µM solution in water/acetonitrile; 1:1 v/v with 0.1 % formic acid) on a Q-Exactive HF Orbitrap (Thermo Scientific) equipped with an electrospray ion source (source voltage 3.5 kV, 275 °C capillary temperature and no sheath gas) via Ultimate 3000 nano UPLC (Dionex) system, with an external calibration (Thermo Scientific). The resolution was 240.000 at m/z = 400 and the mass range was m/z = 160-2000.

To prepare 4-SBD in salt form with tetrafluoroborate as counterion, a suspension of sulfanilic acid (2.00 g, 11.5 mmol, 1.0 eq) in 48% aqueous tetrafluoroboric acid (4 mL, 30.3 mmol, 2.6 eq) was cooled to -5 °C. While stirring a pre-cooled aqueous saturated solution of sodium nitrite (1.20 g, 17.3 mmol, 1.5 eq) was added dropwise to the reaction mixture. During the addition, it became increasingly difficult to maintain stirring which was solved by adding 4 mL of cooled ultrapure water. After this, the reaction mixture was stirred at -5 °C for 5 h after which the product was isolated by filtration. The collected precipitate was washed with ice-cold diethyl ether, ice-cold diethyl ether/methanol (4:1 v/v), ice-cold ethanol and dried overnight under vacuum to yield 4-SBD as a white solid (2.43 g, 9.94 mmol, 77 %). The compound was stored at 4 °C. $^1$H NMR (400 MHz, DCl 0.1 M/DMSO-$d_6$) δ 8.63 (d, $J$ = 8 Hz, 2H), 8.18 (d, $J$ = 8 Hz, 2H). $^{13}$C NMR (100 MHz, DCl 0.1 M/DMSO-$d_6$) δ 156.7, 134.8, 130.2, 118.5. $^{19}$F NMR (376 MHz, DCl 0.1 M) δ -150.6. ESI-HRMS (m/z): $C_6H_5N_2O_3S^+$ [M+] calculated: 185.00, found: 185.00. For all experiments 4-SBD was prepared fresh or used within a timeframe of two months while being stored at 4 °C between experiments.

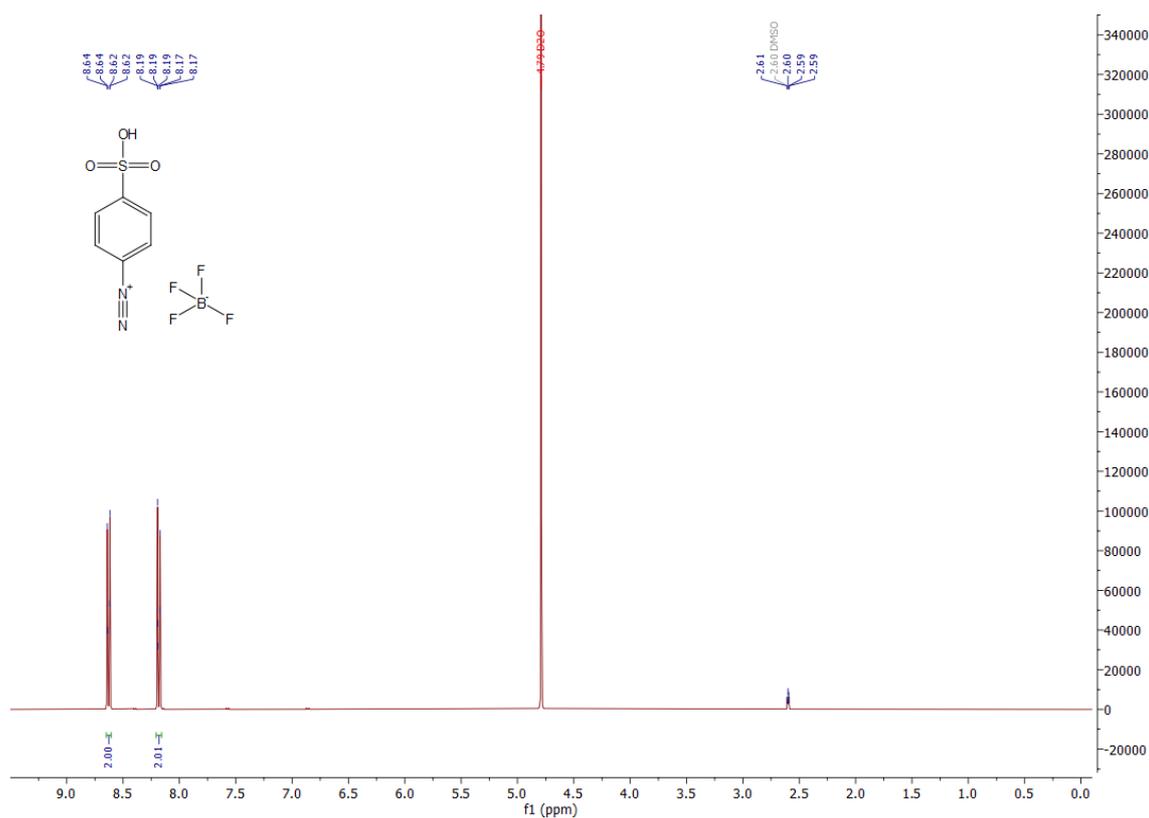

**Fig. S1** $^1$H NMR spectrum of 4-SBD.

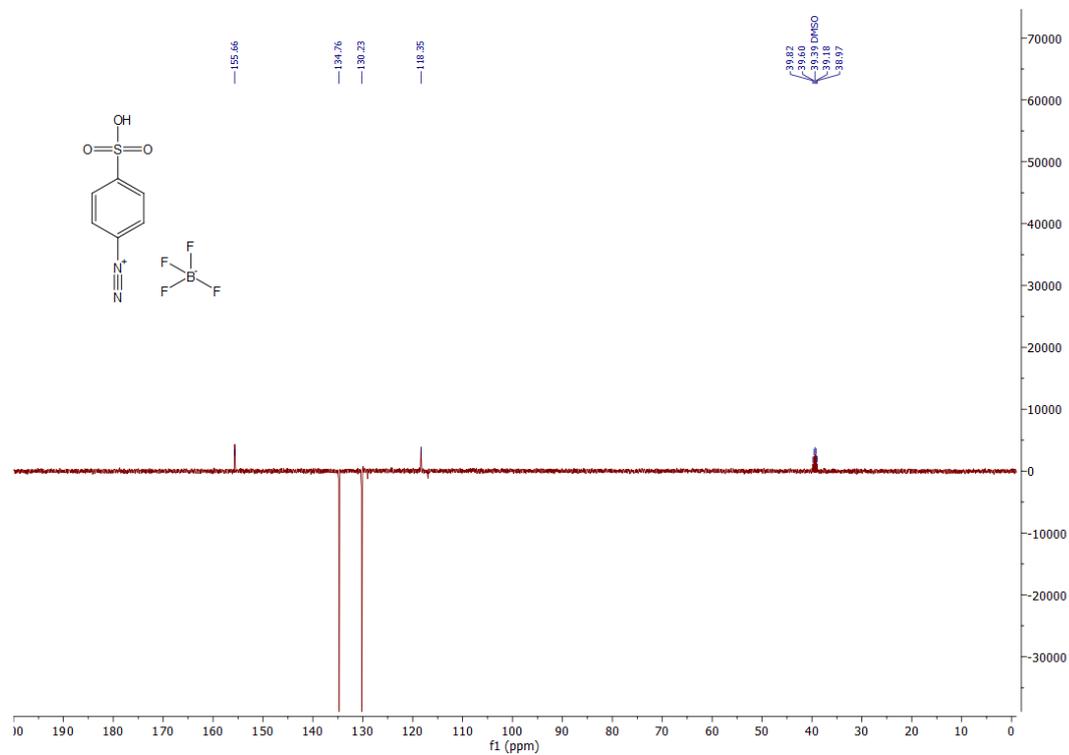

**Fig. S2** $^{13}$C-APT spectrum of 4-SBD.

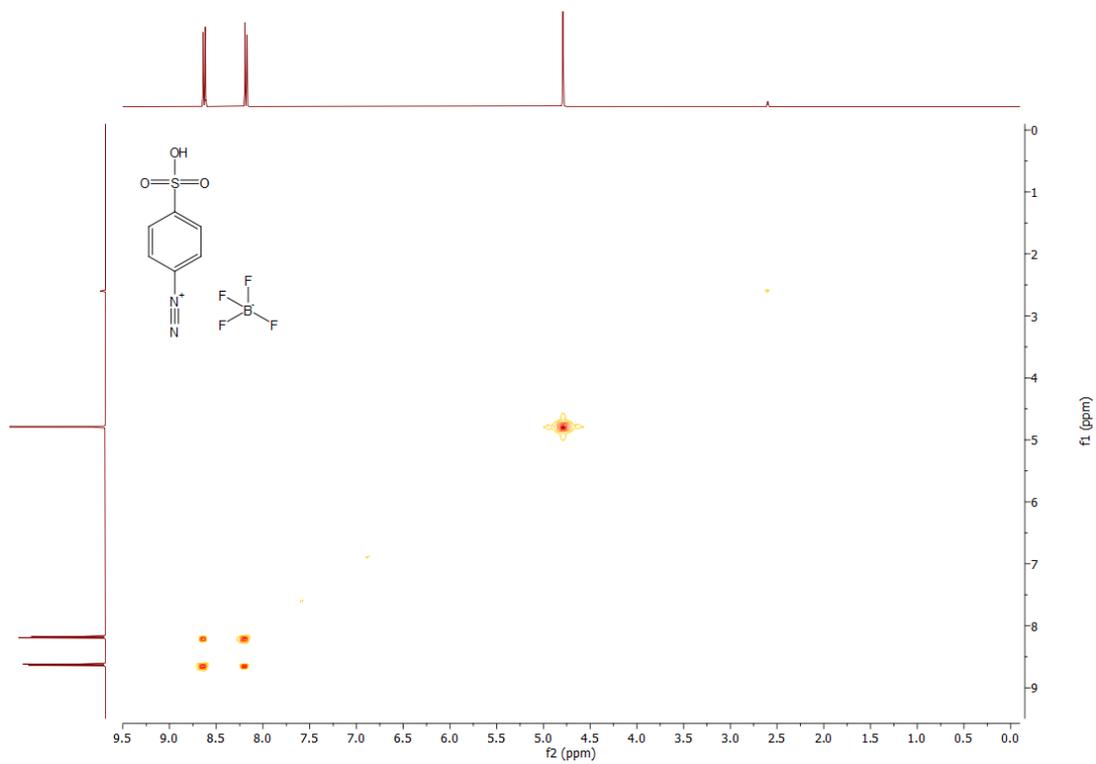

**Fig. S3** COSY spectrum of 4-SBD.

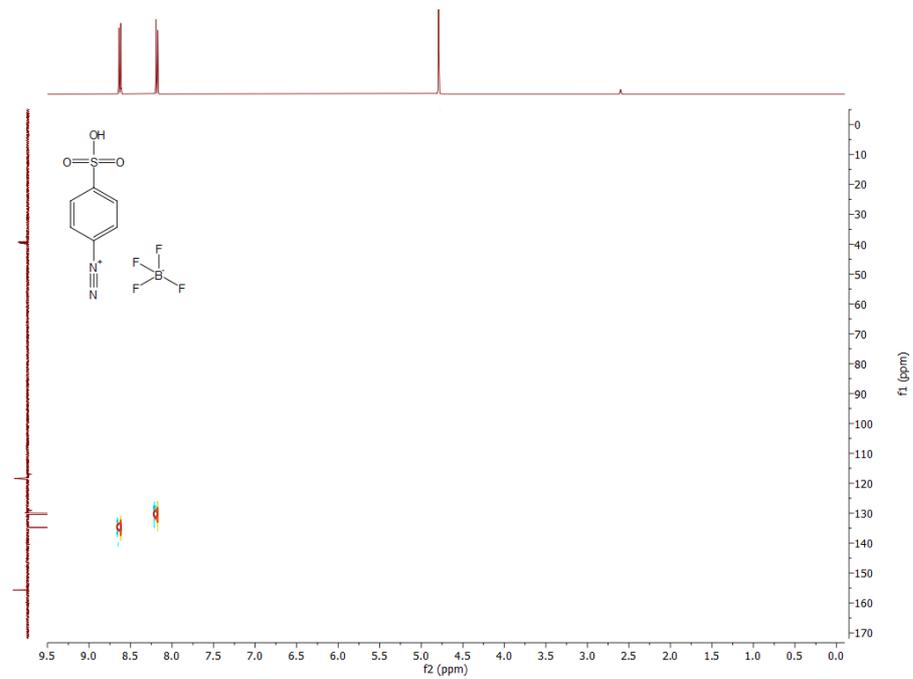

**Fig. S4** HSQC spectrum of 4-SBD.

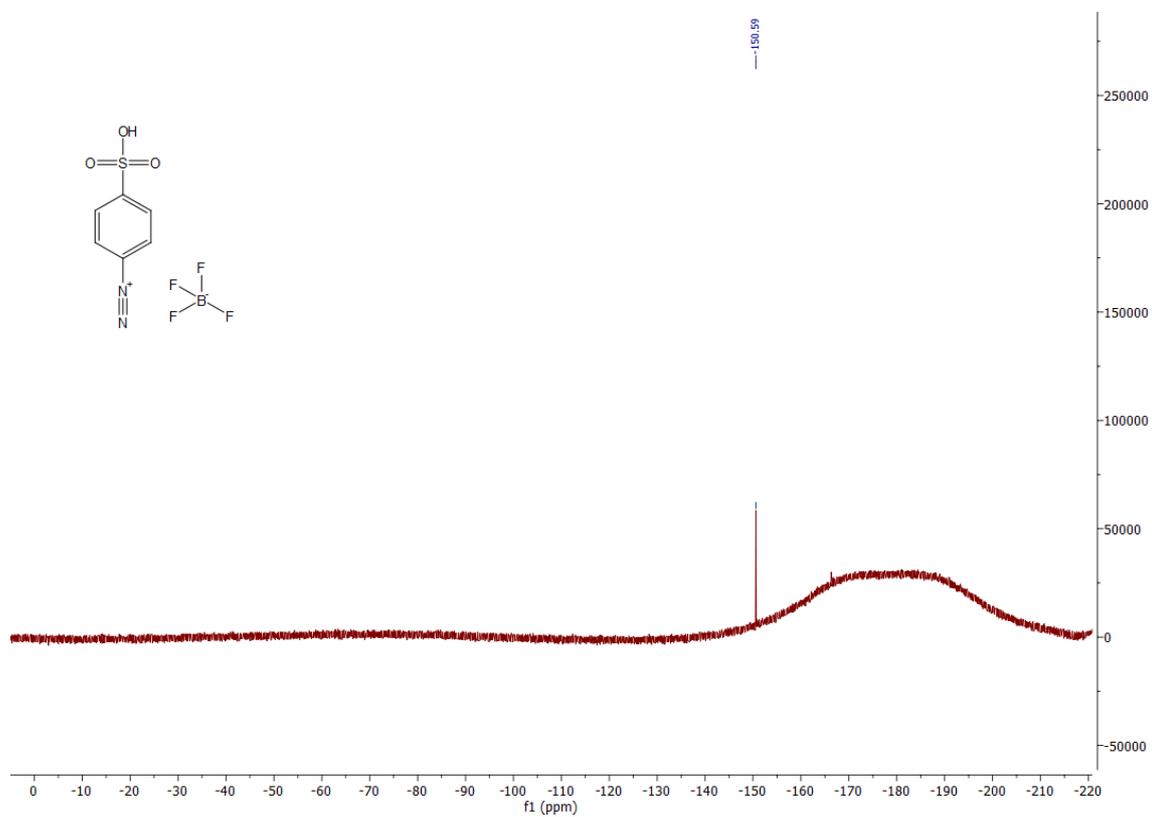

**Fig. S5** $^{19}$F NMR spectrum of 4-SBD.

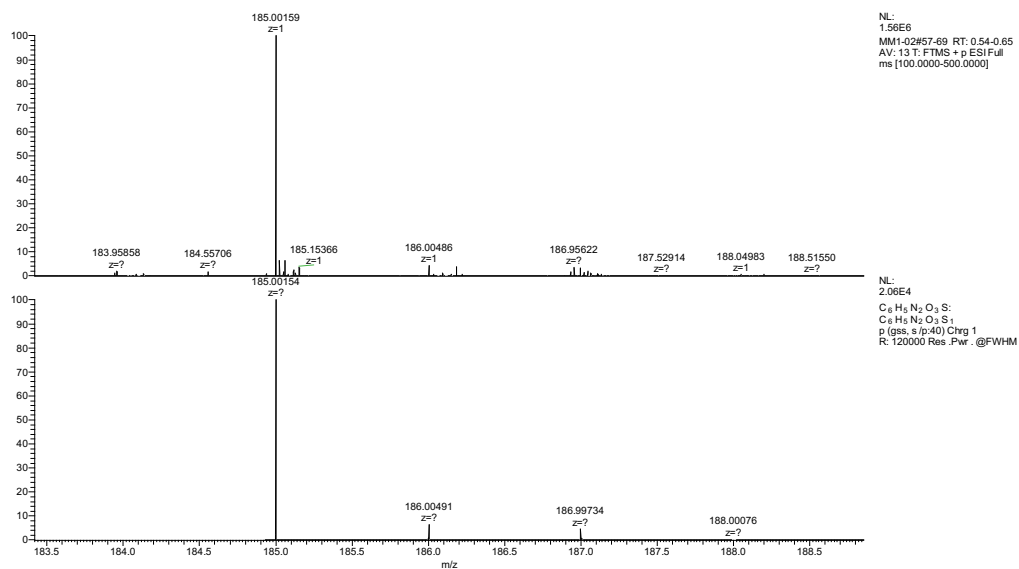

**Fig. S6** HRMS spectrum of 4-SBD.

**Colorimetric assay of 4-SBD reactivity**

To confirm the reactivity of 4-SBD at any time, a solution of 4-SBD (1 mg ml$^{-1}$) in 0.1 M aqueous HCl was mixed with dimethylaniline to obtain methyl orange, a common pH indicator. After mixing, the colorless mixture rapidly turned red or yellow depending on the pH, which is indicative of methyl orange. Mixing a solution of 4-sulfophenolic acid, the major degradation product of 4-SBD, with dimethylaniline did not result in any color change.

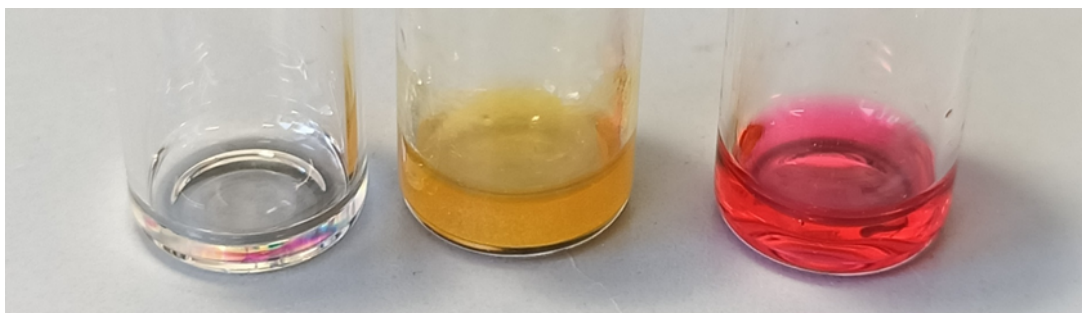

**Fig. S7.** From left to right: aqueous mixtures of dimethylaniline with 4-sulfophenolic acid, dimethylaniline with 4-SBD at pH > 4.4, and dimethylaniline with 4-SBD at pH < 3.1. pH-dependent colorization after mixing dimethylaniline and 4-SBD indicate that methyl orange has been successfully formed.

## 2. Proton (ions) conductance measurements of exfoliated graphene before and after 4-SBD treatment.

**2.1 Ionic current measurements**

The 4-SBD solution (1 mg ml$^{-1}$) was prepared by sonicating 50 mg of the 4-SBD molecule (synthesized in section 1) in 50 ml of aqueous 0.1 M HCl for 5 mins. Unless mentioned otherwise, all the treatment with 4-SBD uses an as-prepared solution of 1 mg ml$^{-1}$. Current-voltage curves of exfoliated graphene were obtained using an Axopatch 200B patch clamp system(*1*). We prepared exfoliated graphene using scotch tape(*2, 3*) and transferred monolayer flakes to a SiN$_x$/Si chip using wedging transfer(*4*). Fig. S8 shows a typical exfoliated graphene flake on SiO$_2$/Si with graphene of different layers marked as 1, 2, and 3. In this study, we only use the area of the monolayer (marked as 1) for the ionic current measurements. Fig. S9 shows exfoliated monolayer graphene used in the fabrication of the devices. All exfoliated monolayer graphene flakes were checked using Raman spectroscopy, which showed a peak intensity ratio of I$_{2D}$/I$_G$ higher than 2. This indicates that the flakes were indeed monolayers. The SiN carrier chips with a circular 1 μm aperture were fabricated by lithography as described in Fig. S10.

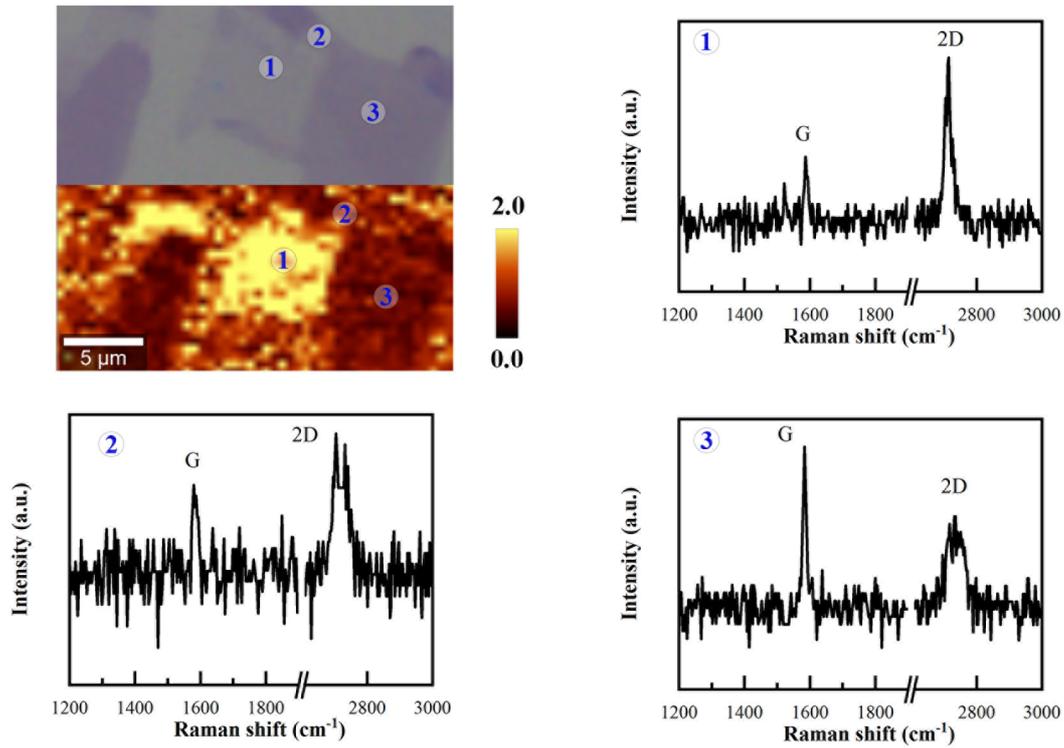

**Fig. S8 Raman characterization of exfoliated graphene.** Optical micrograph of the exfoliated graphene flake on a SiO$_2$/Si with 285nm of thermal oxide and the Raman mapping results of I$_{2D}$/I$_G$ (left-top) with spectra of areas marked as 1, 2, and 3.

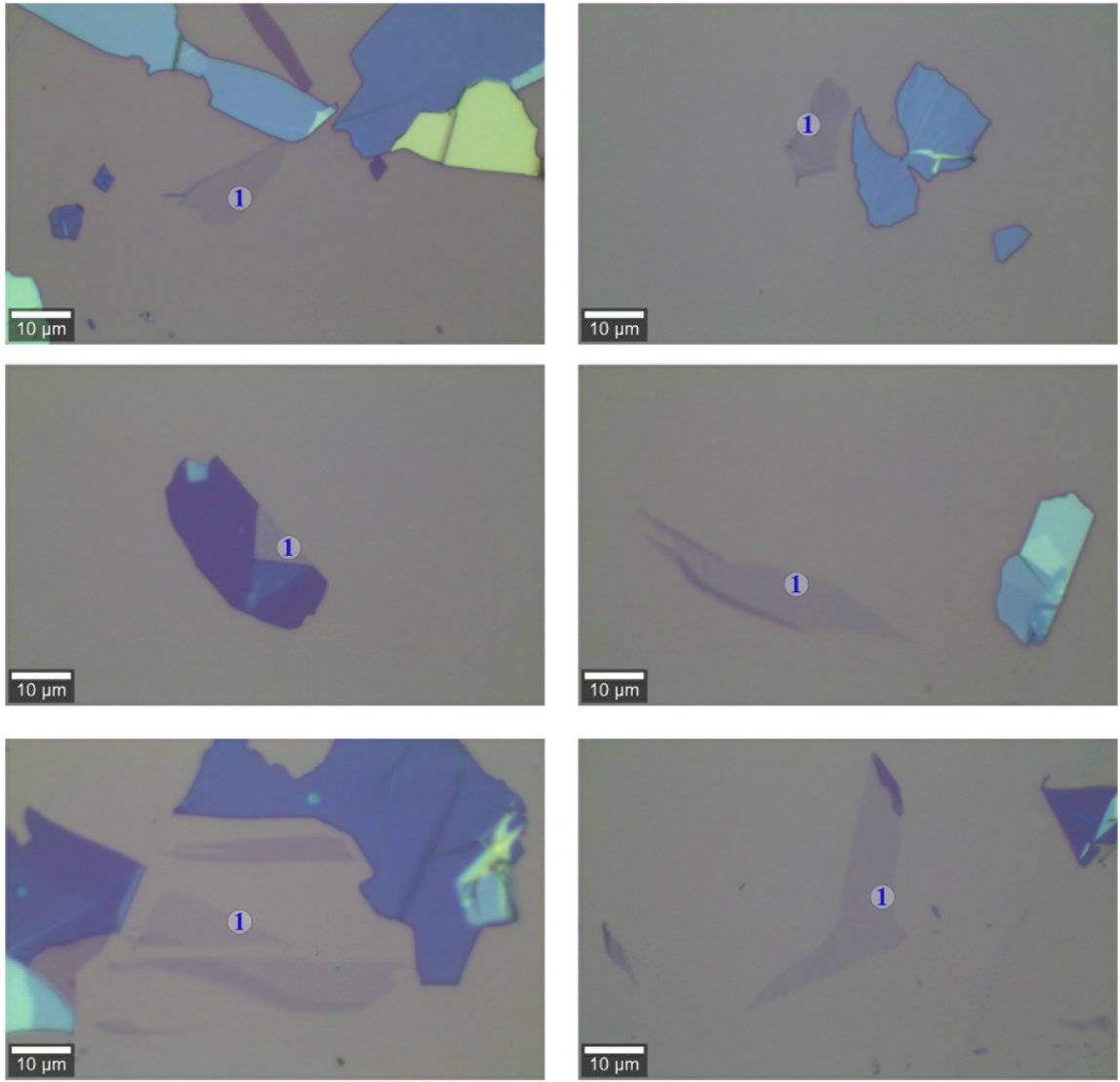

**Fig. S9** Optical micrographs of exfoliated graphene flakes on a $SiO_2$/Si used in the device fabrication.

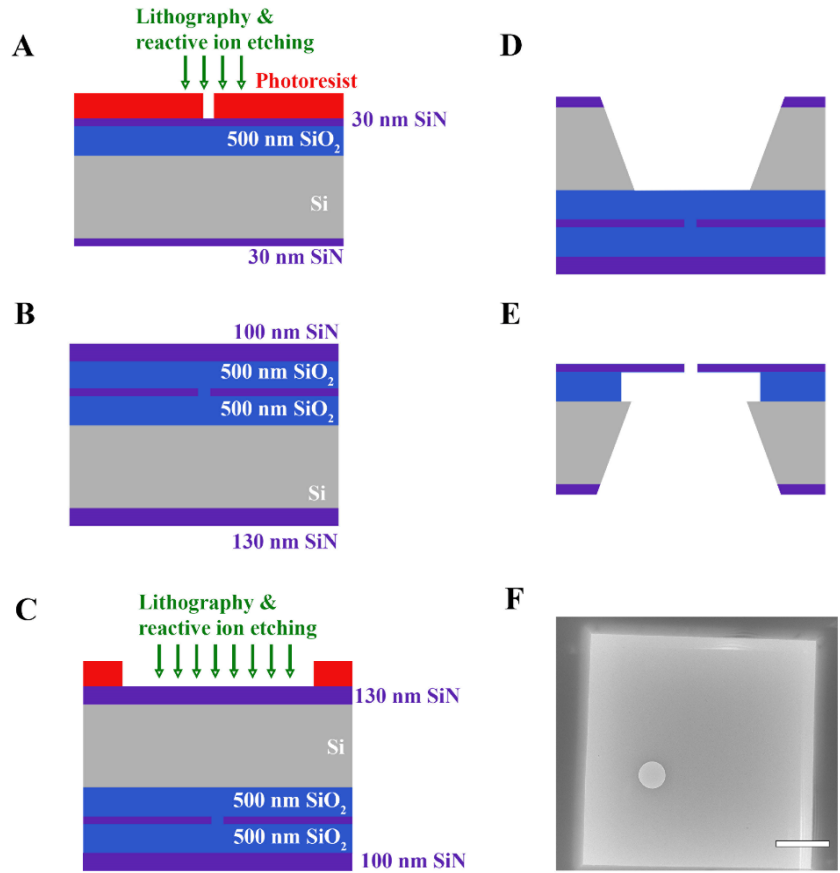

**Fig. S10 Lithographic steps used in the fabrication of the SiN carrier chip.**
**(A)** After the PECVD oxide and the LPCVD membrane nitride layer are deposited, lithography is done to expose a photo-sensitive resist layer with a stepper reticle. The reticle is designed in such a way that the resist is illuminated only with a 1μm in diameter disk. After development, the exposed resist is dissolved. The nitride layer is open and is etched away by RIE etching. **(B)** After stripping the resist there are two sacrificial layers deposited on top – a PECVD oxide layer and an LPCVD nitride layer – to protect the LPCVD membrane nitride layer for step (D). **(C)** On the backside of the wafer, there is again a lithographical step done by using a second reticle. By RIE etching, the nitride is etched away. This forms the window for further KOH etching. **(D)** In a KOH solution, the Si substrate is etched selectively to the lattice of the Si. KOH is etching <1,0,0> planes 100 times faster than <1.1.1> planes. The KOH lands on the backside of the first PECVD oxide layer. **(E)** At the frontside, the last sacrificial LPCVD nitride is stripped by RIE etching. The PECVD oxide layers on both sides of the LPCVD membrane nitride layer are etched in a buffered oxide etch solution. **(F)** SEM micrograph of the resulting chip. Notice that the hole is not exactly in the middle of the window, because of the vicinal angle in the Si substrates. Scale bar: 2 μm.

Upon applying a bias voltage between the two Ag/AgCl electrodes, we measured the ionic current. To monitor the change in ionic conductance during the 4-SBD treatment, we regularly measured the ionic current with the 4-SBD solution refreshed every 24 hours (Fig. S11 showing the exfoliated graphene flake used in device 1). To begin, we flushed the chambers with an ethanol/water mixture (1:1, *v/v*) to remove any gas clogged in the channels. We then flushed the chambers three times with a freshly prepared 4-SBD solution. After the 4-SBD treatment, we flushed the chambers three times with 0.1 M HCl before conducting ionic current measurements for $H^+$ conductance. Next, we repeated the process with 0.1 M KCl for $K^+$ conductance measurements. We repeated the $H^+$ and $K^+$ conductance measurements two additional times to ensure the reproducibility of the ionic conductance for the device after a specific treatment time. Finally, we flushed the chambers again with a 4-SBD solution to continue with another period of 4-SBD treatment. Fig. S14 shows the ionic currents that were measured in device 1 in response to applied voltages at specific treatment times To ensure reproducibility, the same measurement was conducted on a second device. Fig. S12 shows the exfoliated graphene flake used in device 2. After 5 days of 4-SBD, we could still clearly observe the intact monolayer graphene flake covering the aperture of the chip. We had a relatively low success rate in the ionic conductance experiments with exfoliated graphene samples. Utilizing meticulous experimental records, we determined the device success ratio upon commencing formal characterization of proton ionic current through exfoliated graphene. While initial stages saw a higher rate of discarded devices, it's important to note that these failures primarily stem from manipulation factors. Among 17 devices, 2 devices survived the 6 days of experiment particularly after the replacement of solutions over 100 times. 3 devices underwent abrupt increases of current during the 4-SBD treatment, owing to graphene delamination. 12 devices showed no change after several days treatment and present asymmetric I-V curves, indicating the capacitance current is dominant. Fig. S13 shows the delamination of the graphene flake during measurements. Fig. S18 shows the exfoliated graphene flake that was utilized in device 3, which enabled the observation of giant proton conductance through 4-SBD treated graphene for the first time. Fig. S19 shows the I-V curves measured with device 3. It's worth noting, for the differences of device 3 compared to device 1 and device 2, that the 4-SBD used in the experiment was stored at 4°C for six months which means the active 4-SBD component might be lower. Additionally, the SiN chip used was fabricated by directly etching a 60 nm SiN membrane deposited on a Si wafer with the backside etched to create a 1 μm aperture.

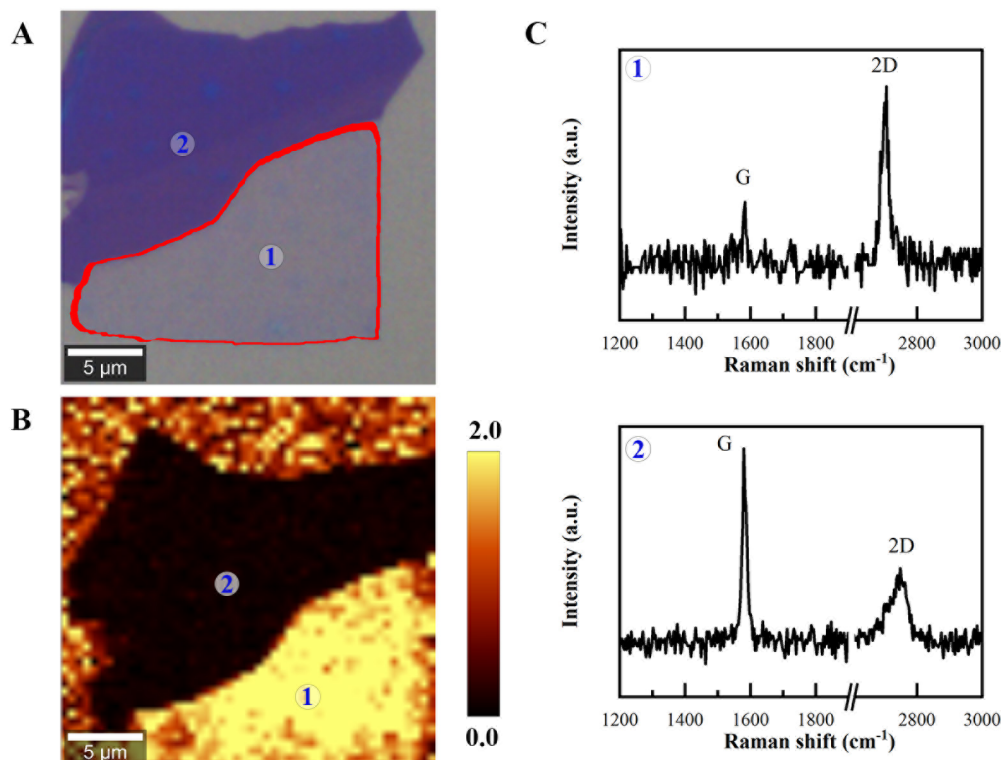

**Fig. S11 Raman characterization of exfoliated graphene used in device 1**. Optical micrograph of the flake on a SiO$_2$/Si with 285nm of thermal oxide **(left-top)** and the Raman mapping results of I$_{2D}$/I$_G$ **(left-bottom)** with spectra of areas marked as 1 and 2 **(right)**.

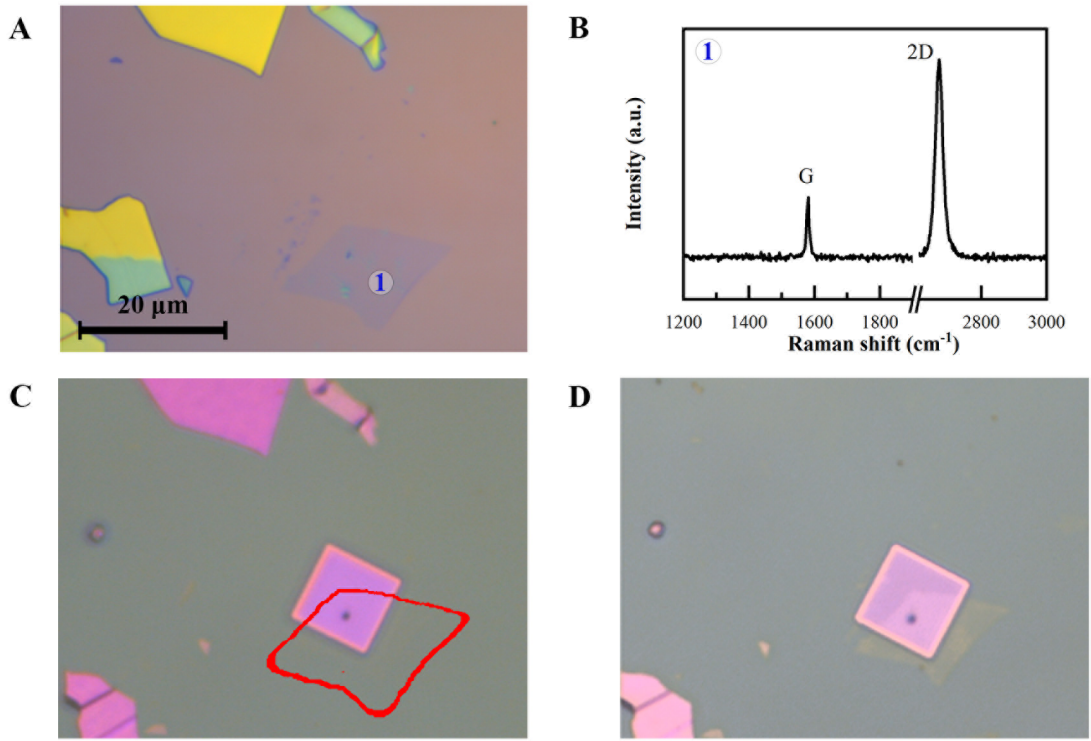

**Fig. S12 Optical and Raman characterization of the exfoliated graphene flake used in device 2 (supplementary information). (A)** Optical micrograph of the flake on a SiO$_2$/Si with 285nm of thermal oxide. **(B)** Raman spectra of the flake (457 nm laser, 1.5 mW). **(C)** Optical micrograph of the flake on the SiN chip after transfer. **(D)** Optical micrograph of the flake on the SiN chip after 5 days of 4-SBD treatment and ionic current measurements.

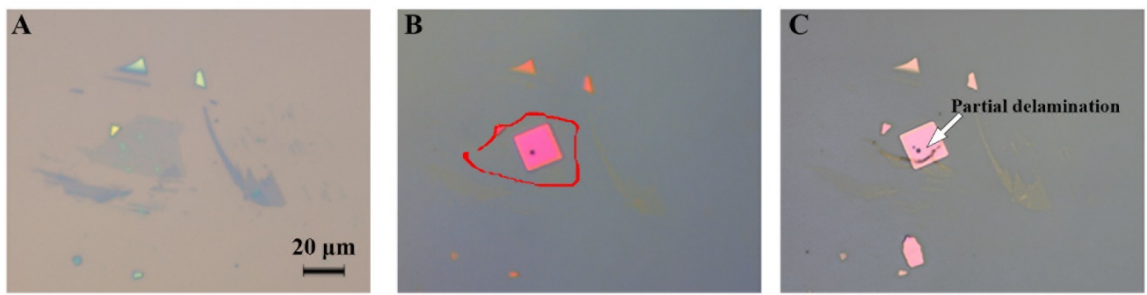

**Fig. S13 Optical characterization of a failed exfoliated graphene device. (A)** Optical

micrograph of the flake on a SiO$_2$/Si with 285nm of thermal oxide. **(B)** Optical micrograph of the flake on the SiN chip after transfer. **(C)** Optical micrograph of the flake on the SiN chip after we observed a large leaking current during measurements (after 2 days of 4-SBD treatment). The flip-over of the graphene flake is visible in this picture, indicating a partial delamination.

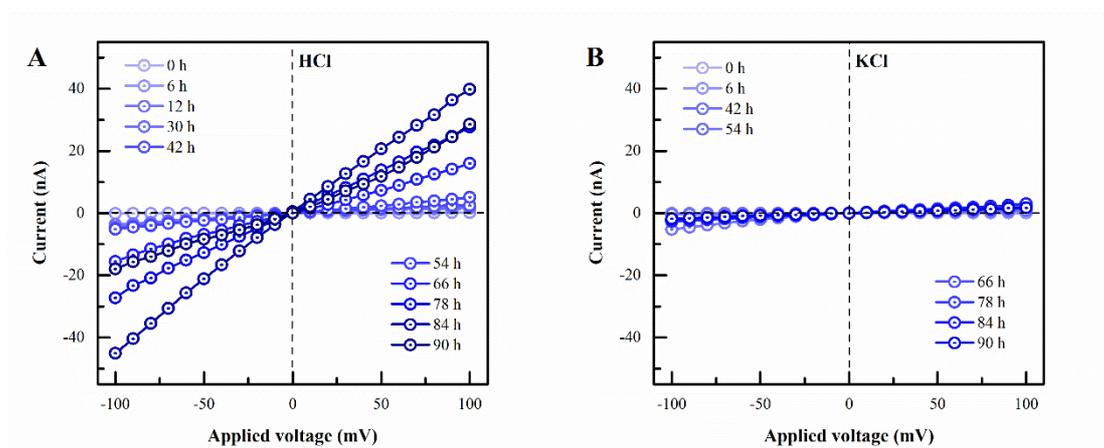

**Fig. S14** I-V curves of exfoliated graphene (device 1) after the specific treatment (1 mg ml$^{-1}$ 4-SBD in 0.1 M HCl) time in 0.1 M HCl (**A**) and 0.1 M KCl (**B**), under ambient conditions.

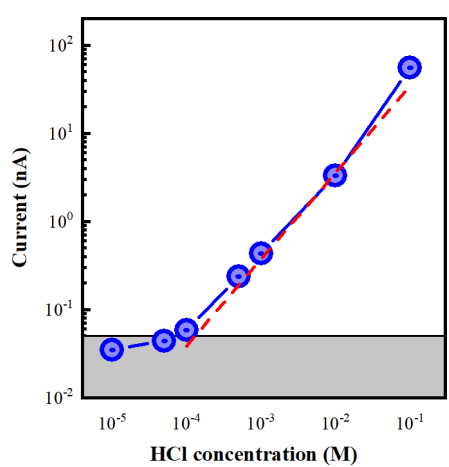

**Fig. S15** Ionic current relationship to the concentration of HCl of device 1 after 84 h treatment. With the concentration of HCl diluted to lower than 10$^{-4}$ M, the relationship

between conductance and concentration is no longer linear.

For ion selectivity measurements, we first determined the proton selectivity over chloride by setting concentration gradients on the two sides of the graphene. As shown in Fig. S16, at the concentration gradient with a factor of 10, the membrane potential is ~56 mV which indicates a perfect cation selectivity(5). As a result, the ionic current observed in the experiment only reflects the movement of cations such as protons and $K^+$. Subsequent ionic current measurements using different chloride electrolytes serve to demonstrate only the transport of cations. For the concentration gradient larger than 50, the membrane potential will no longer fit the linear relationship to the logarithmic concentration ratio. This is consistent with the relationship between conductance and concentration shown in Fig. S15. When the concentration of bulk solution has a comparable amount of charge to the charge fixed on the membrane surface, the influence of the concentration gradient is reduced(6). For the diluted solution, the potential difference between the two sides of the membrane still obeys the electro-osmosis model but the ion distribution near the membrane is not negligible.

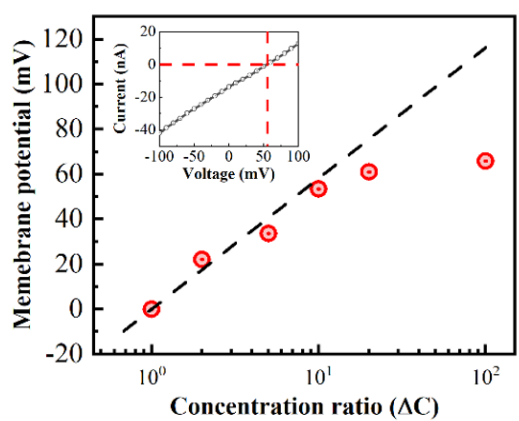

**Fig. S16** Membrane potential for 4-day treated SO$_3^-$-graphene measured by setting concentration gradients of HCl solutions in two chambers (device 1). Inset: I-V curve of 0.1:0.01 M HCl. The dashed line indicates membrane potential with the selectivity of proton to Cl$^-$ based on the Nernst equation. Concentration gradients of 1, 2, 5, 10, 20, and 100 were created using aqueous HCl solutions with concentrations of 0.1 M:0.1 M, 0.1 M:0.05 M, 0.1 M:0.02 M, 0.1 M:0.01 M, 0.1 M:0.005 M, and 0.1 M:0.001 M. These gradients were used to measure the membrane potential of 4-day treated SO$_3^-$-graphene in device 1.

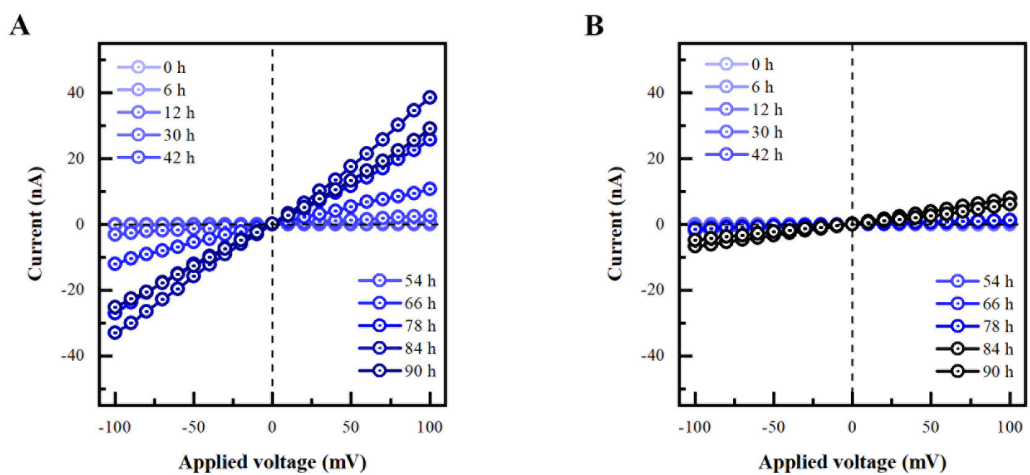

**Fig. S17** I-V curves of exfoliated graphene (device 2) after 4-SBD treatment (1 mg ml$^{-1}$ in 0.1 M HCl) time measured in 0.1 M HCl (**A**) and 0.1 M KCl (**B**), under ambient conditions.

To compare the transport of different cations, chloride salts were normalized with the elimination of electrolyte conductivity differences by:

$$g_i = G_i \frac{\sigma_{HCl}}{\sigma_i} \quad (1)$$

where $g_i$ is the normalized conductance to the proton, $G_i$ is the measured conductance, and $\sigma_i$ is the bulk conductivity of solution i ($\sigma_i$ of 0.1 M solutions at ~20 °C, HCl: 3.70 S m$^{-1}$; CaCl$_2$: 1.59 S m$^{-1}$; CuCl$_2$: 1.75 S m$^{-1}$; FeCl$_3$: 2.55 S m$^{-1}$; KCl: 1.28 S m$^{-1}$).

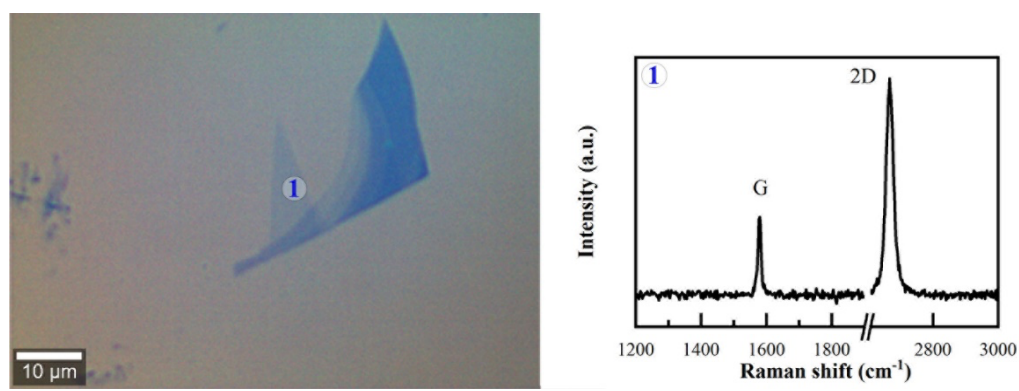

**Fig. S18 Optical and Raman characterization of the exfoliated graphene flake used in device 3**. Optical micrograph of the flake on a SiO$_2$/Si with 285nm of thermal oxide **(left)**. Raman spectrum (457 nm laser, 1.5 mW) of the flake **(right)**.

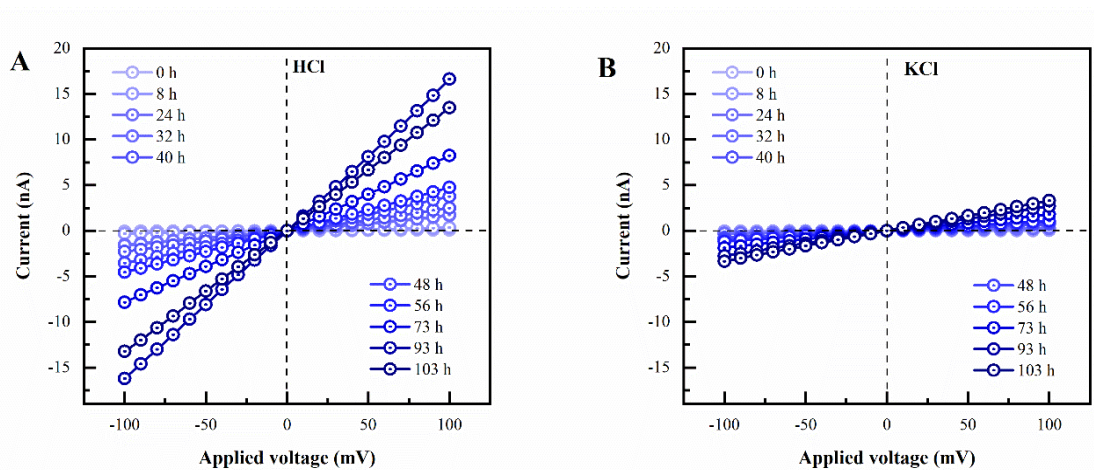

**Fig. S19** I-V curves of exfoliated graphene (device 3) after 4-SBD treatment (1 mg ml$^{-1}$ in 0.1 M HCl) time measured in 0.1 M HCl (**A**) and 0.1 M KCl (**B**), under ambient conditions.

## 2.2 Theoretical studies on proton transport.

**ReaxFF-MD simulations.**

The ReaxFF-MD simulations(*7-9*) are performed using the AMS2021 suite(*10*) and the CHONSSiNaFZr.ff force field(*11*) suitable for describing these systems in a water environment(*12*). We build up the starting system from a 17.22 Å × 17.04 Å graphene layer placed in a cell of 17.22 Å × 17.04 Å × 25.00 Å dimension (Fig. S20). A sulfophenylated $sp^3$ functionalization is added to the graphene basal plane. The system is solvated with 170 water molecules to reach a density of ≈ 1.0 g mL$^{-1}$. Periodic Boundary Conditions (PBC) are applied. ReaxFF-MD equilibrations are performed for 0.5 ns each, with a time step of 0.25 fs and a damping constant of 1000 and 100 for pressure and temperature, respectively. First, an NPT equilibration run is performed at 300 K and 1 atm using the Martyna-Tobias-Klein (MTK) barostat and the Nosé-Hoover chains (NHC) thermostat. Secondly, the system is further equilibrated in the NVT ensemble with the NHC thermostat at 300 K. The dimension of the simulation cell for the NPT equilibration is kept constant along x- and y-the axis (17.22 Å × 17.04 Å parallel to the graphene sheet) and changes only along the z-axis to 24.23 Å. The total charge is -1. After adding a proton to the system in the water bulk, the total charge is equal to 0, and another 0.5 ns NVT equilibration is executed with the NHC thermostat at 300 K. Finally, the production run NVT ReaxFF-MD plus metadynamics(*13, 14*) simulation of 0.5 ns is performed employing the PLUMED plugin(*15*) for the metadynamics, NHC thermostat at 300 K. We employ a well-tempered metadynamics method with width = 0.5 Å, height = 0.1 kJ mol⁻¹ and deposition frequency of Gaussian hills every 100-time steps(*16*). In Fig. S20 the obtained free energy profile and activation Gibbs free energy barrier are shown.

The Collective Variable used for describing the proton traveling from the sulfophenylated sp³ functionalization into the water bulk is defined as the distance from the center of mass of the oxygens of the sulfonic group to the oxygen atom of the hydronium(*17*).

$$L = f(n_G) \left[ \frac{\sum_{i\in\{O_w\}} d_i \exp(\lambda n_i)}{\sum_{i\in\{O_w\}} \exp(\lambda n_i)} \right]$$

where $d_i = |r_i - r_G|$ is the distance between oxygen atom $i$ and the center of mass of the oxygen atoms belonging to the functional group, and $n_G$ is the hydrogen coordination number of the functional group. The switching function $f(n_G)$ included in the Collective Variable is given by:

$$f(n_G) = \frac{1 - \left(\frac{n_G}{n_c}\right)^6}{1 - \left(\frac{n_G}{n_c}\right)^{12}}$$

where the coordination cutoff $n_c$ is a parameter whose value depends on how many excess protons are present in the system.

The switching function $f(n_G)$ is included in the Collective Variable since the distance calculated by the function in square braces would not be meaningful if the functional group is protonated. For instance, at the start of the ReaxFF-MD metadynamics simulation when the functional groups are protonated, the value of the Collective Variable is near zero since $n_G = 1$ ($f(n_G) \approx 0$) and consequently, the Collective Variable is close to zero. The switching function $f(n_G) \approx 1$ resulting in Collective Variable depending on the value in square braces and describing the distance between the center of mass of the oxygens of the sulfonic group to the oxygen atom of the hydronium.

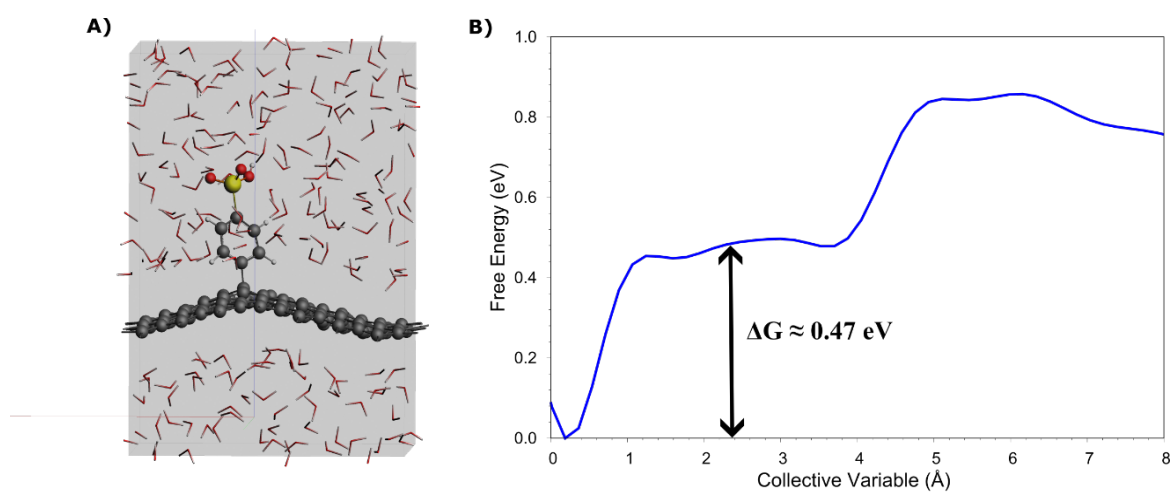

**Fig. S20 A**) Representative ball and stick snapshot of the sulfophenylated sp³ defect functionalization of graphene layer solvated in water, with the simulation box represented in grey. **B**) Free energy profiles (eV) along the Collective Variable (Å) over 0.5 ns of simulation for graphene with sulfophenylated sp³ defect functionalization, blue line. The Gibbs free energy barrier of ΔG ≈ 0.47 eV is indicated by the black arrow.

3. **Single-crystalline graphene growth and characterization.**

Large-area high-quality single-crystalline graphene samples were epitaxially grown on Cu(111) foils via the chemical vapor deposition (CVD) method(*18*). Decimeter scale commercially available polycrystalline Cu foils were placed in a homemade low-pressure CVD system, which is equipped with a 6-inch quartz tube. An asynchronous heating process was conducted and therefore a temperature gradient was then applied to the Cu foils, which promotes the Cu grain growth and leads to the formation of large-area Cu(111) single crystal(*19*). The heating and annealing process was carried out under Ar gas (1000 sccm, 500 Pa), in which trace amounts of oxygen could clean the surface of Cu foils and passivate the active sites of Cu to suppress the adlayer formation and nucleation density. Subsequently, keeping the temperature at 1020 °C, graphene growth proceeded under a gas mixture of $H_2$ (500 sccm) and $CH_4$ (0.8 sccm). Typical SEM images of as-grown graphene on Cu(111) with 30 and 60 mins are shown in Fig. 2B. The well-aligned hexagonal graphene domains indicate the unidirectional orientation of graphene lattice and free of grain boundaries in the seamlessly stitched graphene film.

Transmission electron microscopy (TEM) and selected area electron diffraction (SAED) measurement (JEM-2100, at 200 keV electron energy) were also conducted on the as-obtained graphene samples, which were transferred onto TEM grids (3 mm in diameter) without polymer transfer medium. As shown in Fig. S21, a series of SAED patterns collected on different positions across the graphene sample are with the same orientation, which implies the single-crystal nature of as-grown graphene. Furthermore, the intensity profile of the diffraction pattern along the red dashed line in Fig. S21f indicates that it is monolayer graphene.

We also characterized the transferred graphene on SiO$_2$/Si substrate by using optical microscopy (OM) and Raman spectroscopy (Fig. S22). The high uniformity of optical contrast in OM image (Fig. S22a) and noise-level D-band intensity in Raman spectra (Fig. S22b) confirm the high quality and uniformity of as-grown monolayer graphene (*20, 21*).

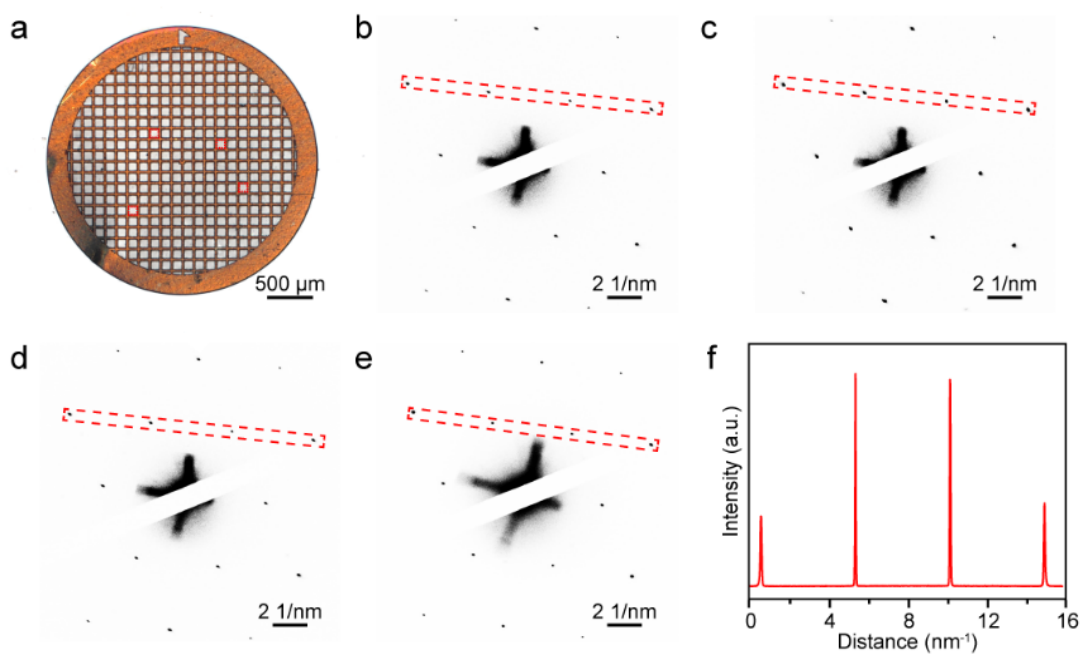

**Fig. S21 Characterization of graphene lattice orientation. (a)** Optical microscope image of the suspended graphene transferred onto a TEM grid. **(b-e)** Typical SAED patterns of the suspended single-crystal graphene measured at the marked red quadrans position in (a). **(f)** The intensity profile of the diffraction pattern along the red dashed line in (e), indicates a typical graphene lattice size.

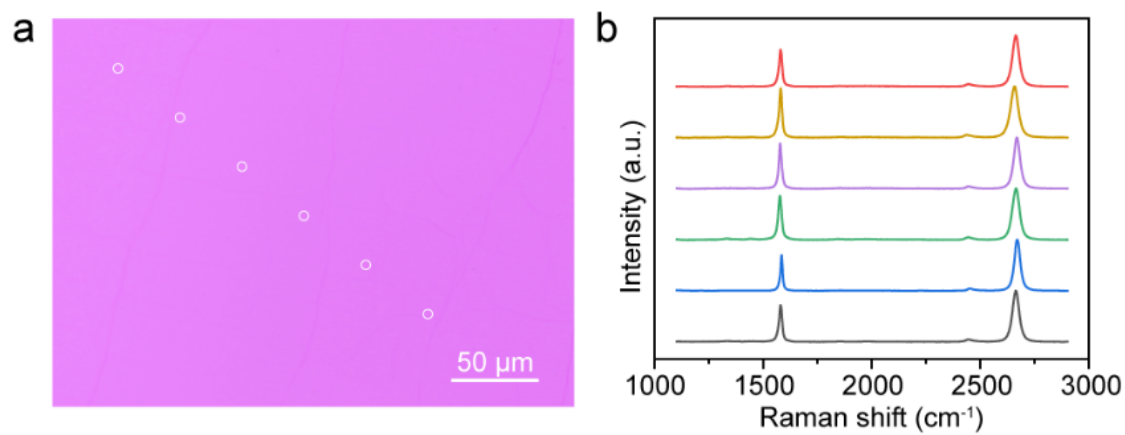

**Fig. S22 Characterization of the single-crystalline graphene film. (a)** Optical microscopy image of monolayer graphene transferred onto a SiO$_2$/Si substrate. **(b)** Representative Raman spectra of the as-transferred graphene measured at the marked position in (a).

4. **Characterizations of single-crystalline CVD graphene after 4-SBD treatment (Raman spectroscopy, XPS, AFM, and HRTEM)**

**4.1 Raman spectroscopy**

Raman spectra were recorded on a WITec confocal Raman spectrometer with a 457 nm laser and a 100× objective under 1.5 mW laser power.

For Nafion/graphene samples, we spin-coated a Nafion solution (~5% D521 from Fuelcellstore) on single-crystalline graphene on a copper foil at 2000 rpm for 1 min followed by baking on a hotplate at 60 °C for 30 mins to fully remove the solvent. Then, the Cu film was etched away by floating the copper/graphene/Nafion stack on an 0.5M aqueous solution of ammonium persulfate (APS, from Sigma-Aldrich) in a glass petri dish. Nafion/graphene was then rinsed by transferring the film onto the water-air meniscus (again floating) in a fresh ultrapure water bath for 20 mins. This procedure was repeated 3 times. For the 4-SBD treatment, Nafion/graphene was floated on the 4-SBD solution (0.1 mg ml$^{-1}$) in a Petri dish for the time indicated. Then, the samples were floated on 0.1 M HCl for 40 mins and rinsed 3 times for 20 mins each time in ultrapure water. Finally, the Nafion/graphene stack was transferred to a SiO$_2$/Si wafer (with 285 nm thermal oxide layer, purchased from Siegert Wafer) before Raman characterization. Fig. S23 shows the mapping of the intensity ratio of the D peak and 2D peak over the G peak from the Raman spectrum indicating the homogeneity of 4-SBD grafting.

For graphene/SiO$_2$ samples without a Nafion layer, CVD graphene was transferred onto a silicon wafer with 285 nm SiO$_2$, using spin-coated poly(methyl methacrylate) (PMMA) as a carrier(22). In this process, ~50 μl PMMA solution (6% in anisole, AR-P 662.06, purchased from All resist GmbH), was drop-casted on ~1.5 cm$^{-2}$ graphene on Cu foil. The PMMA was then dried by spin-coating at 4000 rpm for 1 min. The resulting PMMA/graphene/Cu was baked at 80 °C for 30 mins to remove the solvent. Afterward, PMMA/graphene was obtained by etching the Cu in 0.5M APS and rinsing in ultrapure water as mentioned previously. PMMA/graphene was then transferred to a SiO$_2$/Si wafer (~4 cm$^{-2}$). PMMA/graphene on the wafer was baked at 120 °C for 30 minutes to ensure a good adhesion between the graphene and the SiO$_2$ layer(23) followed by 15 minutes of immersion in acetone to remove PMMA (acetone rinsing was carried after the wafer cooled down to room temperature). To further clean the PMMA residue on the graphene surface, fresh acetone, isopropanol, and ethanol were stepwise flushed using a squeezing bottle through the top surface of the graphene. Then the graphene/wafer was baked at 120 °C for half an hour to further improve the adhesion between the graphene and SiO$_2$ surface. Next, a drop of isopropanol (~5 μl for 1 cm$^{-2}$ of graphene) was added on top of the sample to avoid the direct surface tension force when dipping the sample into a 4-SBD solution. Then the samples were immersed in the 4-SBD solution. Before the Raman characterization, samples were rinsed in ultrapure water several times and dried in air.

Mapping of the intensity ratio (2D peak and G peak, $I_{2D}/I_G$) confirmed that the functionalization is uniform over the single-layer area. While the reactivity of the multi-layer is suppressed, indicated by a very low D peak intensity (Fig. S24). Previously, the hindering of diazonium reactivity by the multilayer was discussed as an effect of the substrate(*24, 25*). Therefore, we assumed that the 4-SBD treatment for the preparation of SO$_3^-$-graphene only works for the monolayer graphene. Fig. S25 shows the relationship between 4-SBD treatment time and Raman spectra for graphene on SiO$_2$/Si. For the defect density, though there were reports to conduct quantitative estimation via $I_D/I_G$ and the size of vacancy, the estimation is based on the vacancy-like defects(*26*). In our samples, no D' peak was not found and this further proved that the sulfophenyl groups were covalently grafted to graphene via sp$^3$ distortion.

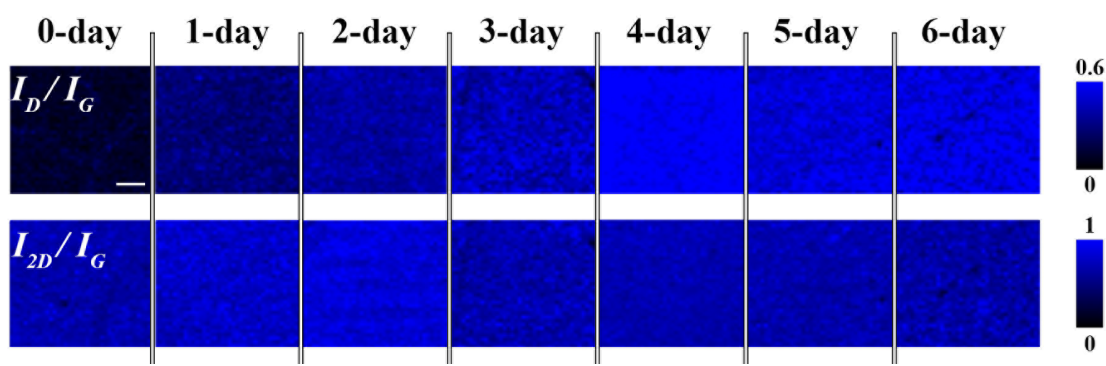

**Fig. S23** Raman mapping of peak intensity ratios, $I_D/I_G$ and $I_{2D}/I_G$, images of Nafion/graphene after floating on SBD solution for 0 to 6 days.

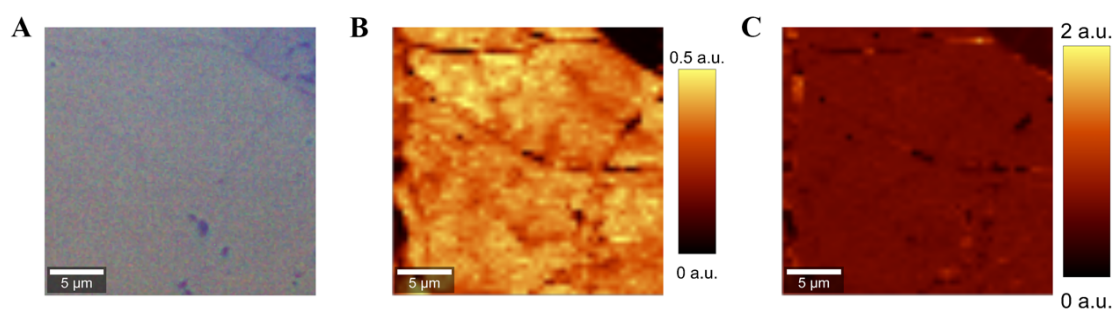

**Fig. S24** Graphene on a SiO$_2$/Si wafer after 4 days of 4-SBD treatment. **(A)** Optical microscopy image. **(B)** Raman mapping of the intensity ratio of D peak over G peak. **(C)** The intensity ratio of 2D peak over G peak.

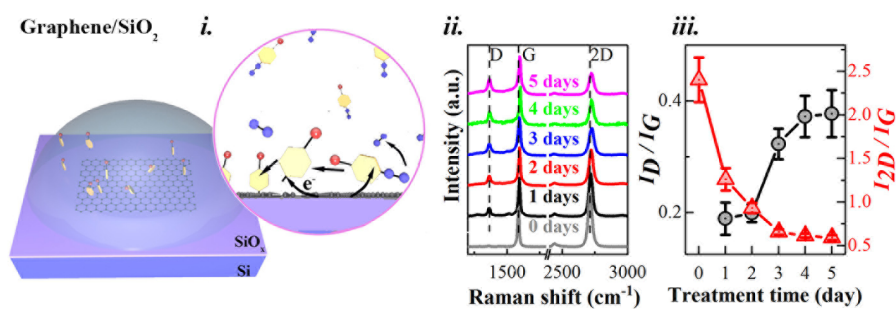

**Fig. S25** i) Illustration of graphene on SiO$_2$/Si reacting with a solution of 4-SBD (1 mg/ml SBD in aqueous 0.1 M HCl). ii) Raw Raman spectra for graphene on a SiO$_2$/Si substrate after 0-6 days of floating incubation with 4-SBD. Raman spectra were recorded over a 10×10 μm area using a 457 nm laser set to 1.5 mW power to avoid any laser-induced damage to the $SO_3^-$-graphene. We normalized the spectra by the intensity of the G peak to facilitate a comparison of the D and 2D peaks. iii) Plot of $I_D/I_G$ and $I_{2D}/I_G$ versus 4-SBD treatment time derived from panel Aii).

**4.2 X-ray photoelectron spectroscopy (XPS)**

Measurements were performed with a Thermo Fisher VG Escalab 250 spectrometer as part of the cluster tool DAISY-SOL. It was equipped with a monochromatic X-ray source (Al Kα= 1486.6 eV) set at 15 mA and 15 kV with a spot size of 650 μm in diameter. The pressure inside the analytical chamber was monitored below $2 \times 10^{-9}$ mbar. Except where otherwise stated, survey spectra were acquired with a pass energy of 50 eV, a step size of 0.1 eV, and a dwell time of 50 ms per measurement point. The detailed scans were acquired with lower pass energy (10 eV) and lower step size (0.05 eV); up to 50 scans were made to increase the signal-to-noise ratio. Metallic foils (Ag, Au, Cu) were cleaned by Ar sputtering (3 kV for 180s) and used for the spectrometer calibration (Fermi edge of Ag at 0.0 eV, Au $4f_{7/2}$ = 84.0 eV, Ag $3d_{5/2}$ = 368.3 eV, Cu $2p_{3/2}$ = 932.7 eV). The Fermi level of the cleaned silver was also used to determine an instrumental resolution of 0.35 eV for XPS (pass energy of 10 eV). After XPS spectra acquisition, they were slightly adjusted (one sample with a shift of 0.4 and the other +/- 0.1 eV) using the C1s $sp^2$ peak at binding energy (BE) of 284.5 eV. For measuring the 4-sulfonatebenzendiazonium (4-SBD) powder on foil, a flood gun was required to bring the adventitious carbon state as close as possible to 284.8 eV; despite that, charge the correction of the spectra to lower BE ($\approx$1.8 eV) were made afterward (Fig. S26).

Interestingly, for the graphene samples immersed in solutions containing 1 and 0.1 mg/mL of SBD, the sp$^3$C/sp$^2$C ratios increase with the concentration of the solution and with the content of SO$_3^-$ detected at the near surface of the graphene samples. A 4-SBD treatment with 10 times less concentrated solution (0.1 mg ml$^{-1}$) leads to a reduction of the XPS peak intensities demonstrating that there are 1.5-3.7 times fewer SO$_3^-$ groups grafted to the graphene basal plane (Table S2). Besides, the two additional C states are assigned to C-N (BE$\approx$ 286.7 eV) and O-C-O (BE$\approx$289.1 eV)(*27*). The small amount of sp$^3$ content inside the pristine sample is assigned to unavoidable C-C contamination of the samples during the polymer-assisted transfer procedure.

For the quantitative chemical analysis and the binding energy (BE) shifts, spectra were analyzed with CasaXPS (version 2.3.19PR1.0) and the Scofied relative sensitivity factor (R.S.F)(*28*). Spectra were fitted by first, subtracting spectral background using a Shirley-type function, using weighted least-squares method and model curves (Voigt functions of 70% Gaussian and 30% Lorentzian for all the peaks and an asymmetric Lorentzian "LA(1.2,2.5,5)" for the sp$^2$ carbon. The fitting methodology has been inspired by the work of Biesinger(*29*). Full half-width maxima (fwhm) of the sp$^2$ C and sp$^3$ C were constrained between 0.4-0.8 eV and 0.9-1.3 eV, respectively. The fwhm of the C-N and O-C-O peaks were constrained to be equal to the one of the sp$^3$ C +/- 0.05 eV. The N=N, NH$_2$, and NO$_x$ peaks were also constrained to have the same fwhm +/- 0.05 eV. The position of the C-N and the O-C-O peaks were, respectively, constrained to be at 1.3-1.7 eV and 3.8-4.3 eV higher energy than the sp$^3$ C peak.

In the N1s core level spectra, three states were identified at roughly 400.3 eV which is assigned to the azo-N=N- of the SBD molecule, or to adsorbed N$_2$ as it is the case for the pristine sample, 402 eV which refers to -NH$_2$ or -NH$_3^+$ probably formed after oligomerization, and 406.1 eV indicates oxidized nitrogen(*30*). -NH$_2$ and oxidized nitrogen are unexpected indicating unknown side reactions besides radical coupling and azo coupling. The peak in the S2p spectrum centered at 168 eV corresponds to -SO$_3^-$ (*31*), which was also observed for the powder SBD precursor (Fig. S26C).

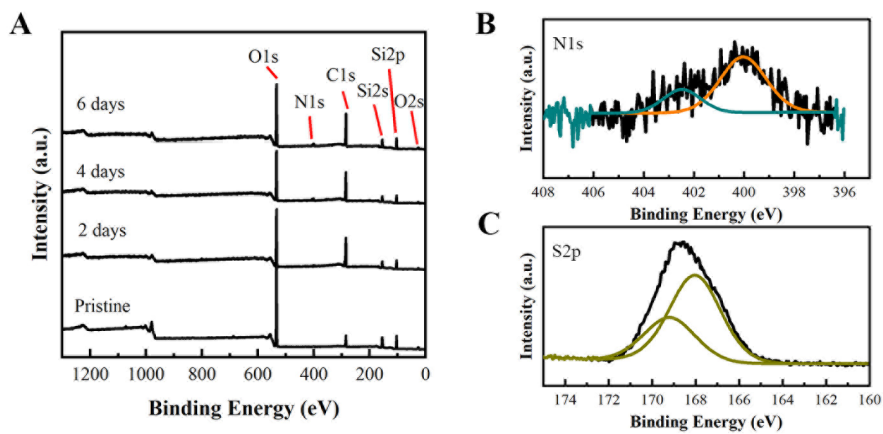

**Fig. S26 (A)** Survey spectra of pristine graphene and graphene after 2 days, 4 days, and 6 days 4-SBD treatment. **(B)** N1s spectrum and **(C)** S2p spectrum of 4-SBD powder.

**Table S1:**

| Duration of treatment | 4-SBD concentration (mg/mL) | Peaks relative concentration (%) | $sp^3C/sp^2C$ ratio |
|---|---|---|---|

|  |  | S2p | C1s | | | | N1s | | | |
|---|---|---|---|---|---|---|---|---|---|---|
|  |  | SO$_3^-$ | sp$^2$ | sp$^3$ | C-N | O-C-O | N=N | NH$_2$ | NO$_x$ |  |
| pristine |  | 0 | 69.9 | 13.0 | 9.2 | 3.3 | 3.9[a] | 0.7[a] | 0 | 0.19 |
| 2 days |  | 2.6 | 36.7 | 36.3 | 13.1 | 9.2 | 1.2 | 0.6 | 0.4 | 1.00 |
| 4 days | 1 | 5.2 | 55.8 | 19.7 | 10.6 | 3.2 | 3.1 | 0.6 | 1.9 | 0.35 |
| 6 days |  | 6.3 | 58.9 | 17.7 | 7.6 | 2.9 | 3.3 | 2.0 | 1.5 | 0.30 |
| 2 days | 0.1 | 1.7 | 39.4 | 34.0 | 13.9 | 7.5 | 2.4 | 0.7 | 0.5 | 0.86 |
| 6 days |  | 1.7 | 39.5 | 33.3 | 13.8 | 10.2 | 0.9 | 0.4 | 0.3 | 0.84 |

[a] Nitrogen content in the pristine sample is assigned to adsorbed $N_2$.

### 4.3 Atomic force microscopy (AFM)

AFM images were recorded using a JPK NanoWizard Ultra Speed microscope and the obtained data was processed using the JPK SPM Data Processing software. All experiments were performed using a silicon probe (Olympus, OMCL-AC240R3) with a nominal resonance frequency of 70 kHz. The images were all scanned and recorded (with a resolution of 1024x1024 pixels) in intermittent contact mode in the air at room temperature. For each incubation time, three individual regions were recorded and the surface roughness was calculated from the combined three regions. The AFM characterization was conducted with graphene on the TEM grid to allow the characterization over free-standing graphene, which is prepared by the following steps. A droplet of isopropanol was added to the TEM grid (carbon film on top), and a piece of flat graphene/Cu film was gently put on the droplet (graphene side on bottom). Then, the sample was dried at room temperature. After the drying out of isopropanol, the sample was flipped over and floated on 0.5 M APS solution to fully etch away Cu film. Then the sample was floated on ultrapure water for 20 mins and this procedure was repeated for 3 times. The as-prepared sample was floated on 0.1 M HCl to preclude the influence of water and HCl on the graphene morphology (shown in Fig. S27 & 28). Then the samples were treated with 4-SBD (1 mg ml$^{-1}$ in 0.1 M HCl solution) in the same manner mentioned above for a specific time, followed by the rinse of 0.1 M HCl after treatment.

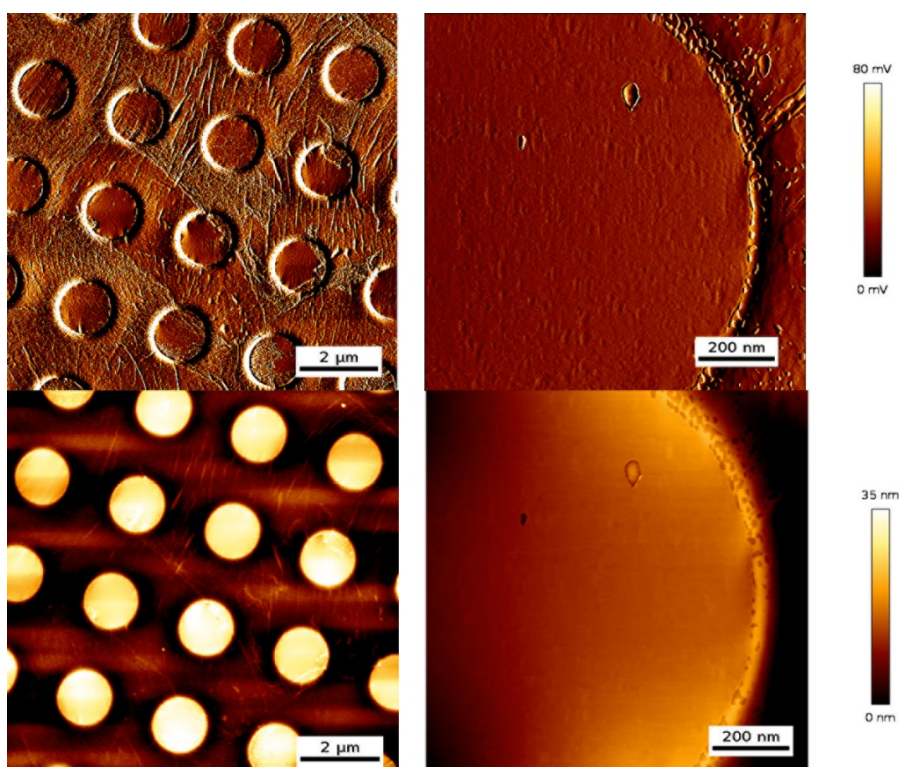

**Fig. S27** AFM images of CVD graphene after floating on 0.1 M HCl solution for 6 days.

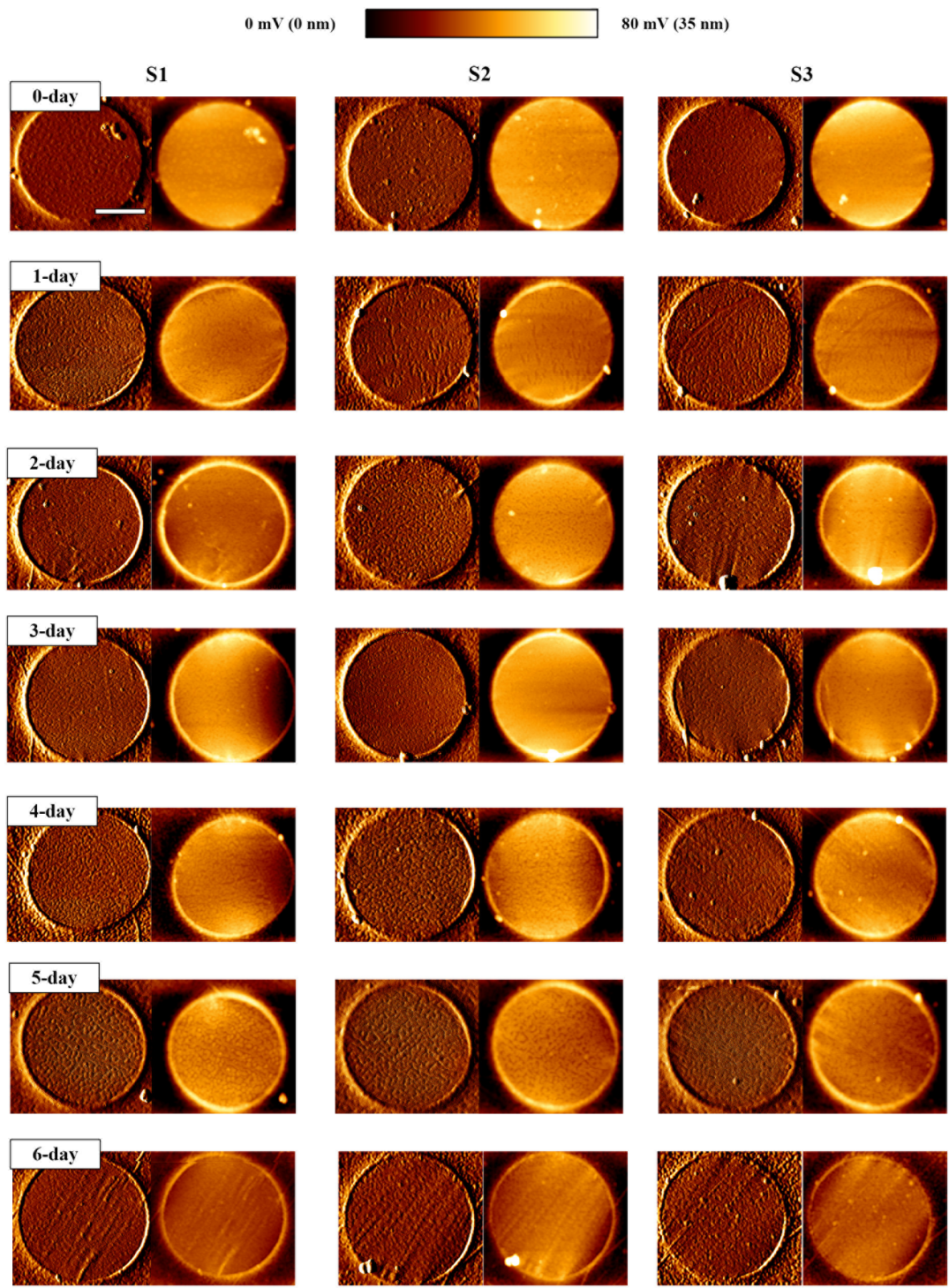

**Fig. S28 AFM images of graphene samples free-standing on TEM grids.** For each of treatment time, three random positions were selected for the AFM mapping. The left image of each set is the error image and the right image is the height image. Scale bar: 500 nm.

**4.4 High-resolution transmission electron microscopy (HRTEM) characterization.**

For the consistency of samples, the samples for HRTEM characterization are prepared in the same manner as samples for AFM characterization. HRTEM images were acquired with the image-side CC/CS-corrected SALVE microscope operated at 80 kV with a resolution of 76 pm (SALVE: Sub-Angstrom Low-Voltage Electron Microscopy). Data acquisition was conducted on a Ceta CMOS camera. We conducted HRTEM characterization of $SO_3^-$-graphene with all three batches of 4-SBD used in this work (Fig. S29 for batch 1, Fig. S30 for batch 2, and Fig. S31 for batch 3). From all these observations, we did not find any vacancy-like defects generated after 4-SBD treatment.

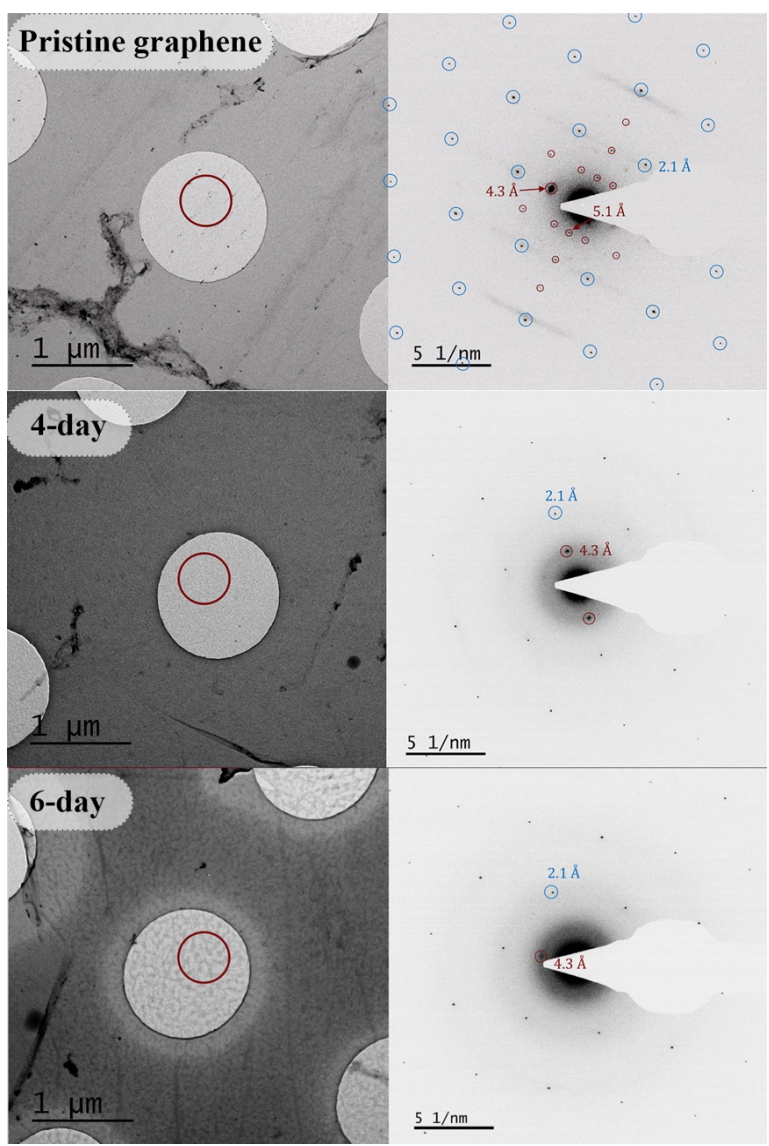

**Fig. S29** TEM image with diffraction pattern of pristine CVD graphene and $SO_3^-$-graphene after 4 days and 6 days 4-SBD (**batch 1**, 1 mg ml$^{-1}$ in 0.1 M HCl) treatment.

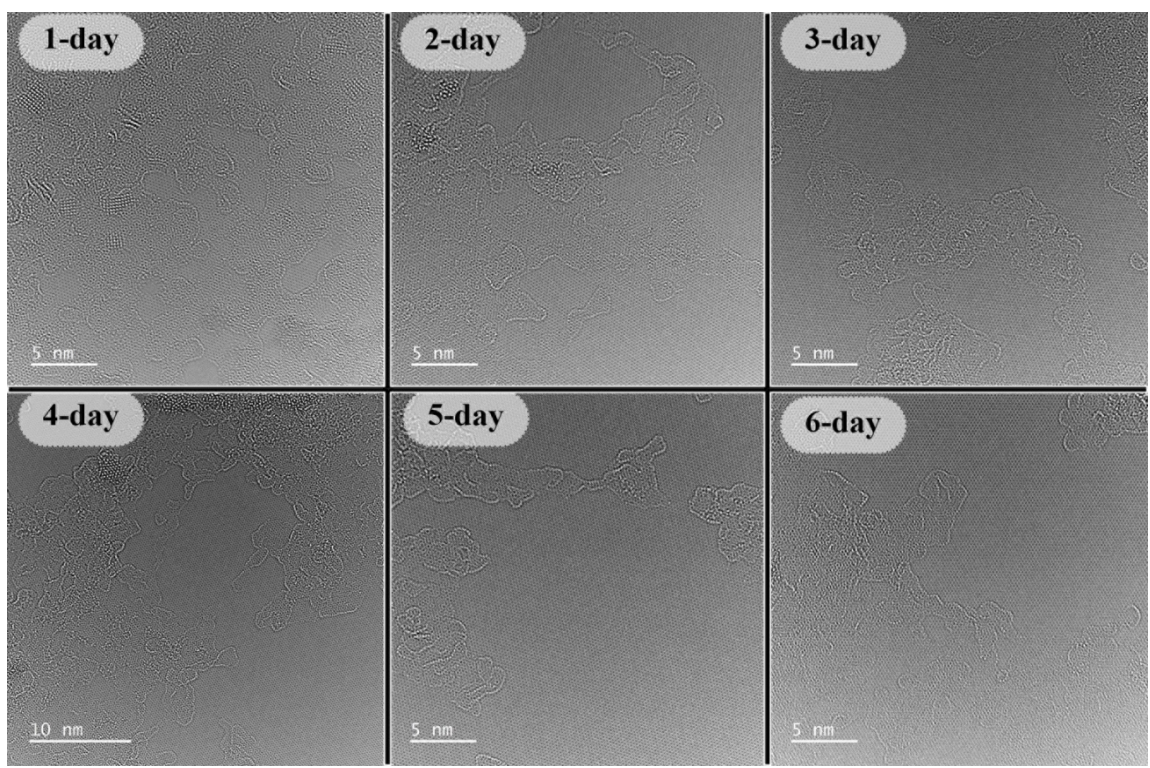

**Fig. S30** HRTEM images of CVD graphene after 1 to 6 days of 4-SBD (**batch 2**, 1 mg ml$^{-1}$ in 0.1 M HCl) treatment.

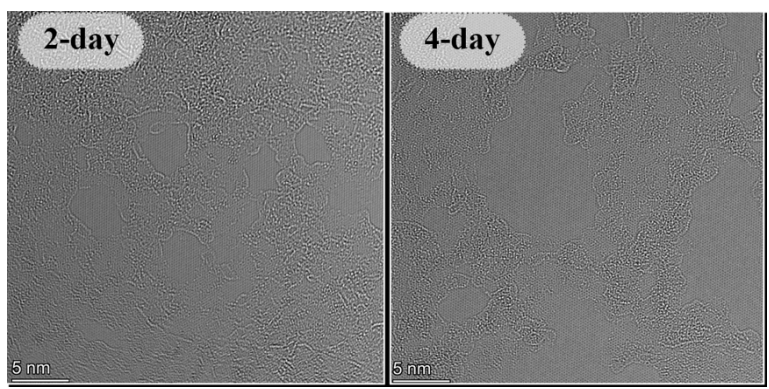

**Fig. S31** HRTEM images of CVD graphene after 2 and 4 days of 4-SBD (**batch 3**, 1 mg ml$^{-1}$ in 0.1 M HCl) treatment.

## 5. Membrane electrode assembly (MEA) fabrication and direct methanol fuel cell (DMFC) operation

### 5.1 MEA fabrication

A single-crystalline CVD graphene membrane was used in a DMFC application, where cracks or defects were minimized during the chemical vapor deposition process. Graphene-based membranes were prepared with a porous polycarbonate (PC) membrane as a support (Nuclepore track-etch membrane with a pore diameter of 3 μm, Whatman), and a proton-conducting ionomer (Nafion dispersion D521, Fuelcellstore) was used as a stabilizer to prevent direct contact between the graphene lattice and electrodes.

To prepare the membrane, CVD graphene on Cu (1.2 by 1.2 cm²) was first spin-coated with the Nafion solution (D521) at 2000 rpm for 1 minute and then baked at 80°C for 30 minutes on a hot plate. The Nafion/graphene/Cu was then placed on a piece of PC membrane, and the Cu was etched by floating the Nafion/graphene/Cu/PC sample on a 0.5 M APS solution in a petri dish. The resulting Nafion/graphene/PC composite membrane was rinsed and stored in 0.1 M HCl before being fixed between two PTFE gasket sheets using soft replica rubber (Reprorubber Thin Pour, Flexbar Machine Corp.). A Nafion solution was drop-casted twice onto the backside of the PC support, followed by baking at 80°C for 30 minutes each time (to act as a mechanical contact buffer during assembly of the fuel cell and following the compression from the electrodes). The anode, with a Pt/Ru loading rate of 4 mg cm$^{-2}$, and the cathode, with a Pt loading rate of 4 mg cm$^{-2}$ (Fuelcellstore), were then put on the sides of the MEA before the measurements.

Fig. S32 depicts the assembly steps.

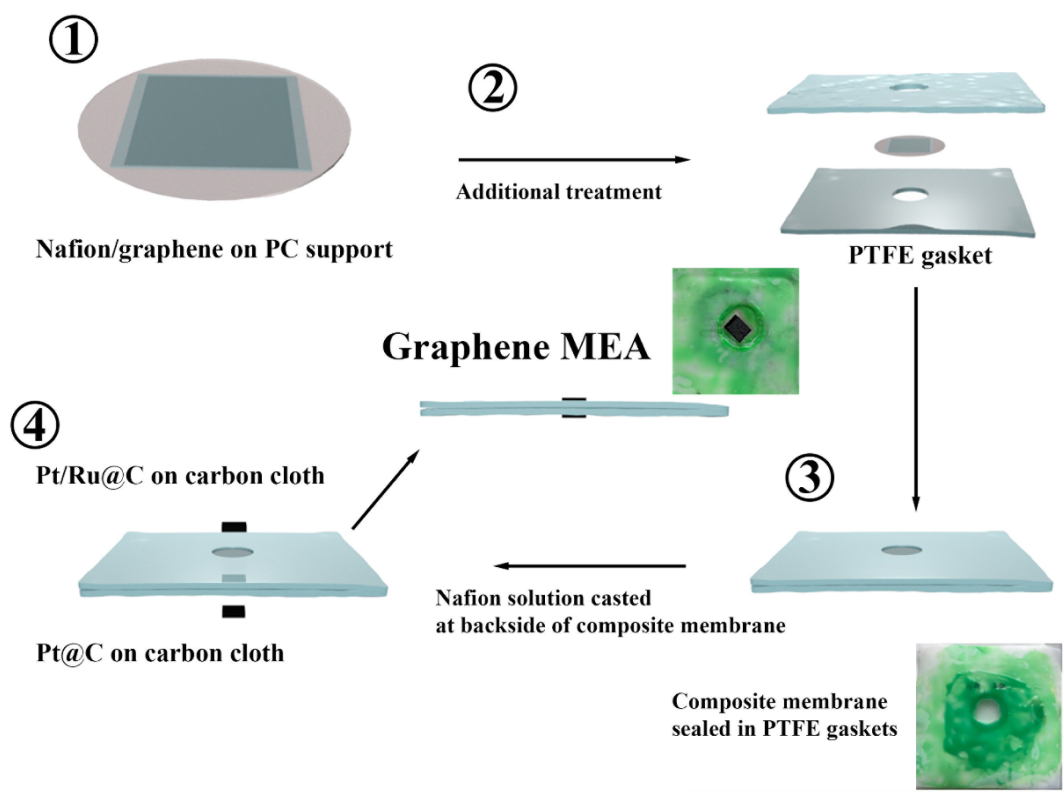

**Fig. S32** Graphene MEA preparation. The samples were floated directly on the 4-SBD solution and subsequently transferred to a 0.1 M HCl solution before use. Next, the sample was fixed between two PTFE gaskets with a circular window of 1 cm using soft replica rubber, ensuring that only the area covered by the graphene sample was exposed. Finally, the electrodes were placed directly on both sides of the assembly before the measurements were taken in the DMFCs.

**5.2 DMFC operation**

All electrochemistry measurements were conducted using a potentiostat (PGSTAT204) equipped with an electrochemical impedance spectroscopy model (FRA32M) and operated using the Nova software.

DMFC measurements were carried out using an electrolyzer from Dioxide Materials, with a titanium anode plate and a cathode made from 904 L stainless steel. The cells were operated using a methanol/water mixture at a flow rate of 30 ml min$^{-1}$ passing through the anode. To ensure the reliability of all measurements, all cells were started with 5 rounds of CV scanning from 0 to 1.5 V at a speed of 10 mV s$^{-1}$, and cathode starvation was performed by floating $N_2$ gas instead of $O_2$, followed by 20 minutes of running the cell at 200 mV. The open circuit voltage was determined by fixing the cell current to 0 and recording it after 5 minutes with no voltage variation. After the cell was fully stabilized and the current variation (under 200 mV) was less than 10 μA min$^{-1}$, the polarization curves were obtained by linear sweep from open circuit voltage to short circuit voltage at a 10 mV s$^{-1}$ scan rate. The resistance was determined by reading the intercept from the Nyquist plot of the electrochemical impedance spectrum, which was measured with frequencies from 1 KHz to 0.1 Hz and 10 mV amplitude under an open circuit. Data were acquired using three independent samples for reproducibility check, and the error bars indicate the maximum and minimum derivation instead of standard derivations for directly showing the experimental results. In addition to using 1 M and 5 M methanol, we attempted to increase the fuel concentration to 10 M methanol. However, this resulted in significant dehydration of the Nafion layer, causing a substantial leakage of methanol and ultimately leading to the destruction of the MEA.

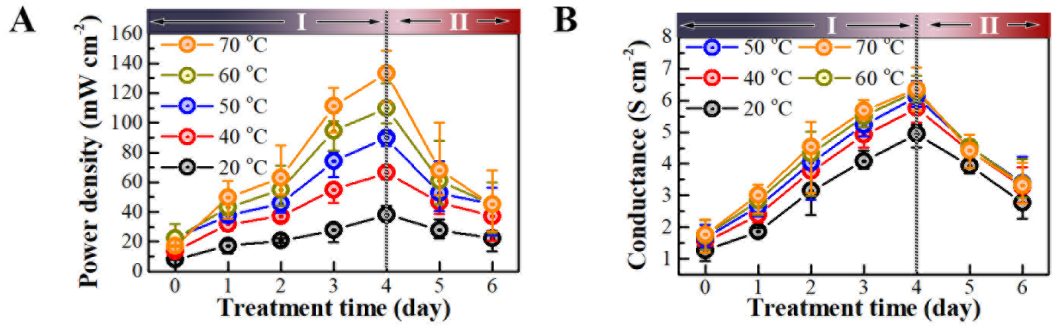

**Fig. S33 (A)** Maximum power density of DMFCs as a function of SBD treatment time for $SO_3^-$-graphene membrane under operating temperature from 20 °C to 70 °C with 1 M methanol/water as fuel. **(B)** Membrane conductance of DMFCs with 1 M methanol and $SO_3^-$-graphene membranes after treatment time from 0 to 6 days under operation temperatures from 20-70 °C.

## 6. Methanol crossover and fuel efficiency estimation

The performance of a DMFC is influenced by methanol crossover in three major ways: it can poison cathode catalysts, reduce methanol efficiency, and decrease cell voltage due to a mixed potential at the cathode. To assess the membrane's properties, we investigated the methanol crossover behavior of an $SO_3^-$-graphene membrane. Methanol crossover rate ($J_m$) is caused by three major effects: gradient diffusion resulting from concentration differences, electro-osmotic drag, and hydraulic pressure based on liquid pressure, which can be expressed as follows:

$$J_m = D \frac{\Delta C}{\delta_m} + n \frac{I}{F} + \frac{K_m \rho \Delta P}{\mu M_{H_2O} \delta_m} = \alpha \frac{I}{F} \qquad (2)$$

Where $D$ is the diffusivity of water in the membrane, $\delta_m$ is the membrane thickness, $\Delta C$ is the concentration difference, $n$ is the electro-osmotic drag coefficient, $I$ is the current density, $F$ is Faraday's constant, $K_m$ is the permeability through the membrane, $\rho$ is the density of water, $\Delta P$ is the difference of liquid pressure, $\mu$ is the viscosity of the liquid, $M_{H_2O}$ is the molecular weight of water, α is the overall coefficient(*32*). With the correlation between the working current and methanol current, we could estimate the methanol crossover current under different working currents by:

$$I_{crossover} = I_{crossover,limit}\left(1 - \frac{I}{I_{shortcircuit}}\right) \quad (3)$$

where $I_{crossover}$ is the methanol crossover current, $I_{crossover,limit}$ is the limit current obtained from crossover current measurements, $I_{shortcircuit}$ is the short circuit current of DMFC and $I$ is the working current(*33, 34*). The ratio of $I_{crossover, limit}$ and $I_{shortcircuit}$ decides the slope of eq. 3 which is in positive correlation with the overall crossover coefficient, α. To further compare the fuel efficiency among samples, we also converted the methanol crossover current information by(*35*):

$$\eta_{fuel} = \frac{I}{I+I_{crossover}} \quad (6)$$

and calculated the fuel efficiency with $I$ of a DMFC measured under 200 mV and $I_{crossover}$ calculated from eq.3 under 200 mV. We observed that the fuel efficiency at 200 mV was due to the maximum power output being achieved around that voltage and the DMFC operating at an optimal state.

Therefore, we characterized the methanol crossover by oxidizing the methanol at the cathode. The methanol crossover current was measured by linear sweep voltammetry from 0 to 1.2 V against the cathode and anode at a scan rate of 5 mV s$^{-1}$ with a methanol flow rate of 30 ml min$^{-1}$ at the anode and nitrogen flow at the cathode. The limiting current is taken as an indicator of the methanol crossover rate.

Fig. S34A displays the I-V curves of a DMFC (red, cathode fed with oxygen) and oxidizing methanol crossover from the anode to the cathode (dark, cathode fed with nitrogen). The crossover current ($I_{crossover}$) represents the amount of methanol that has diffused from the anode to the cathode through the membrane. It is possible to estimate the diffusion efficiency by comparing $I_{crossover}$ with the short circuit current ($I_{shortcircuit}$) using the expression $I_{crossover}$ / $I_{shortcircuit}$(*32*). For the pristine graphene membrane, an increase in temperature from 20 to 70 °C results in a significant rise in methanol crossover, indicated by the ratio of Icrossover / Ishortcircuit increasing from 0.44 to 1.21. Remarkably, after functionalizing graphene with 4-SBD for 2-4 days, $I_{crossover}$ / $I_{shortcircuit}$ becomes constant over the entire range of operating temperatures (Fig. S34B), suggesting that the temperature has a more significant impact on activity than the methanol crossover rate. Additionally, the negligible variation in methanol crossover for samples treated with 2 to 4 days of 4-SBD at different temperatures indicates that the selective pathway dominates. For samples with 5 and 6 days of treatment, the variation with different temperatures is more significant since the selective pathway is partially blocked by the oligomerization of the diazonium compound. The different methanol crossover rates also explain the temperature dependence of the cell performance, as mentioned above. Fuel efficiency at 200mV ($\eta_{200mV}$), which is close to the potential at maximum power density, is a practical Parameter to evaluate the fuel cell performance(*35*). This efficiency was obtained from the methanol crossover current and the working state cell current (Fig. S34C). For the 4-day $SO_3^-$-graphene at 60 °C, $\eta_{200mV}$ reaches ~79% at a current density of 457.5 mA cm$^{-2}$.

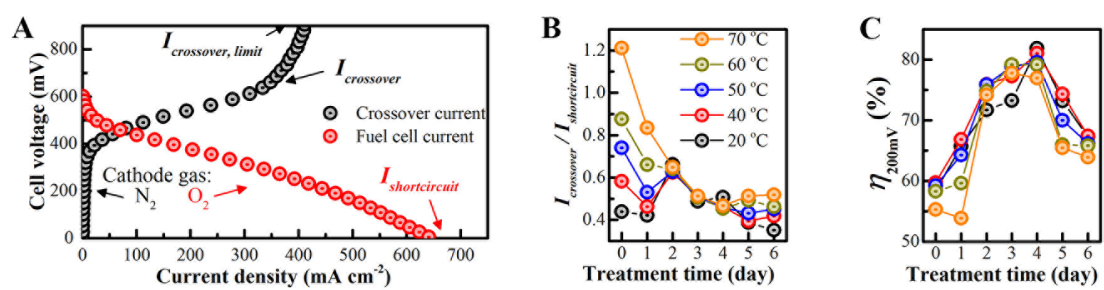

**Fig. S34 Methanol crossover DMFCs with graphene and $SO_3^-$-graphene at different temperatures**. **(A)** Methanol crossover (black) vs. fuel cell current was measured on 4-day $SO_3^-$-graphene (red) at 60 °C with 1M methanol as fuel. In the methanol crossover I-V curve, the diffusion-limited current ($I_{crossover,\ limit}$ plateau region intercepted with x-axis), and in the fuel cell I-V curve, the short-circuited (0 V output) is the maximum current of the DMFC ($I_{shortcircuit}$). **(B)** The ratio of limiting methanol crossover current over short-circuit cell current ($I_{crossover,\ limit}\ /\ I_{shortcircuit}$) as a function of treatment time and under operating temperature from 20 to 70 °C. **(C)** Fuel efficiency at an output voltage of 200mV under operating temperature from 20 to 70 °C (the same color code as B).

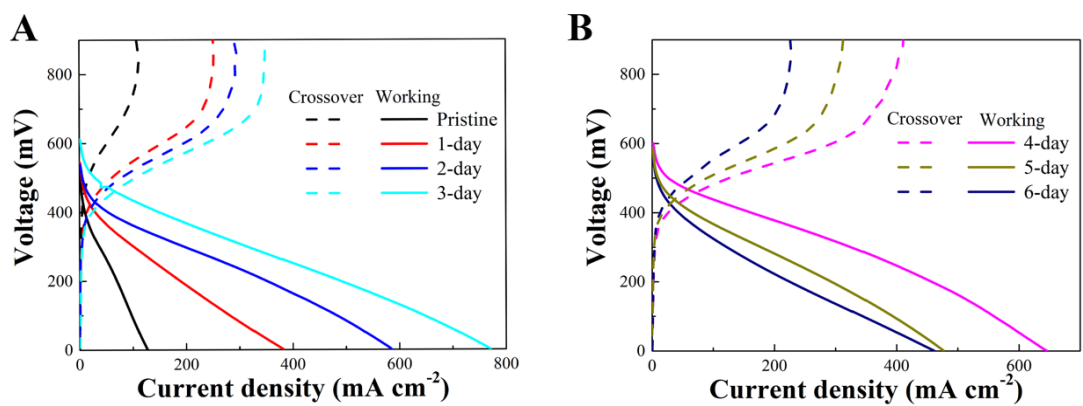

**Fig. S35** Methanol crossover current vs. working current under 60 °C with 1 M methanol/water passing through the anode. **(A)** Pristine to 3-day treatment. **(B)** 4 to 6 days treatment.

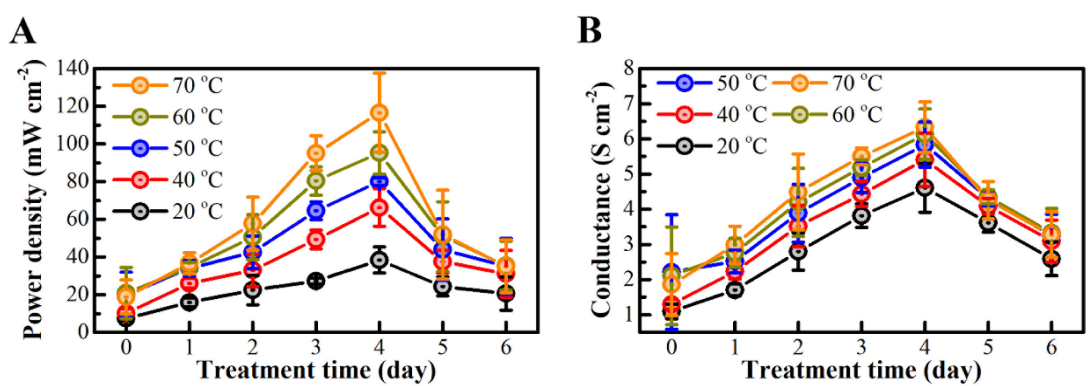

**Fig. S36** Maximum power density **(A)** and conductance **(B)** of samples with different treatment time obtained, measured in DMFCs with 5 M methanol.

**Table S2** Maximum power density of samples with different SBD treatment time under 5 M methanol and operating temperatures ranging from 20 °C (room temperature, rt) to 70 °C. Δ is the standard deviation.

| Treatment time (day) | Max. P (mW cm$^{-2}$), rt | Δ (mW cm$^{-2}$) | Max. P (mW cm$^{-2}$), 40 °C | Δ (mW cm$^{-2}$) | Max. P (mW cm$^{-2}$), 50 °C | Δ (mW cm$^{-2}$) | Max. P (mW cm$^{-2}$) under 60 °C | Δ (mW cm$^{-2}$) | Max. P (mW cm$^{-2}$), 70 °C | Δ (mW cm$^{-2}$) |
|---|---|---|---|---|---|---|---|---|---|---|
| 0 | 8.0 | 0.2 | 13.7 | 4.2 | 22.2 | 9.6 | 25.8 | 15.4 | 17.0 | 0.8 |
| 1 | 17.4 | 4.7 | 31.8 | 3.7 | 37.4 | 3.3 | 42.9 | 4.6 | 49.7 | 13.2 |
| 2 | 21.1 | 1.3 | 38.9 | 2.7 | 48.8 | 8.7 | 60.9 | 10.4 | 70.9 | 17.4 |
| 3 | 27.8 | 6.8 | 54.9 | 7.7 | 74.2 | 9.5 | 94.4 | 11.2 | 111.4 | 15.9 |
| 4 | 38.2 | 5.3 | 66.5 | 4.2 | 89.7 | 4.7 | 109.8 | 14.8 | 133.0 | 17.5 |
| 5 | 28.0 | 6.3 | 46.6 | 12.3 | 55.7 | 16.1 | 61.3 | 23.1 | 64.4 | 31.6 |
| 6 | 22.0 | 7.3 | 37.1 | 14.2 | 45.3 | 18.4 | 44.9 | 17.3 | 45.2 | 21.2 |